\documentclass[aps,twocolumn,prd,reprint,preprintnumbers,superscriptaddress,nofootinbib,floatfix]{revtex4-2}
\RequirePackage[l2tabu, orthodox]{nag}

\usepackage{microtype} 
\usepackage{amsmath,amstext, amssymb, amsthm, amsfonts, slashed}
\usepackage{physics}
\usepackage{mathtools}
\usepackage{graphicx}
\usepackage{tikz}
\usepackage{booktabs}
\usepackage{dcolumn}
\usepackage{bm}
\usepackage{multirow}
\usepackage{dsfont}
\usepackage[normalem]{ulem}
\usepackage{footmisc}
\usepackage[colorlinks=true,urlcolor=blue,linkcolor=blue,citecolor=blue,hypertexnames]{hyperref}
\usepackage[nameinlink]{cleveref}
\Crefname{appsec}{Appendix}{Appendices}
\usepackage{simplewick}
\usepackage{float}
\usepackage{placeins}
\usepackage{lipsum,afterpage}
\usepackage[compat=1.1.0]{tikz-feynman}
\hypersetup{colorlinks=true, linkcolor=blue, citecolor=blue, urlcolor=blue, linktoc=all, linktocpage=true}

\usetikzlibrary{arrows.meta}
\definecolor{proca}{HTML}{6565DE}

\def\s0#1#2{\mbox{\small{$ \frac{#1}{#2} $}}}
\def\0#1#2{\frac{#1}{#2}}

\graphicspath{{./figures/}{./}}

\begin{document}

\preprint{RIKEN-iTHEMS-Report-26}

\title{Cosmology from asymptotically safe Proca theories}

\author{Carlos~Pastor-Marcos}
\affiliation{Institut für Theoretische Physik, Universität Heidelberg, Philosophenweg 16, 69120 Heidelberg, Germany}

\author{Lavinia~Heisenberg}
\affiliation{Institut für Theoretische Physik, Universität Heidelberg, Philosophenweg 16, 69120 Heidelberg, Germany}

\author{\'Alvaro~Pastor-Guti\'errez}
\affiliation{RIKEN Center for Interdisciplinary Theoretical and Mathematical Sciences (iTHEMS), RIKEN, Wako 351-0198, Japan}

\author{Jan~M.~Pawlowski}
\affiliation{Institut für Theoretische Physik, Universität Heidelberg, Philosophenweg 16, 69120 Heidelberg, Germany}
\affiliation{ExtreMe Matter Institute EMMI, GSI Helmholtzzentrum für Schwerionenforschung mbH, Planckstr.\ 1, 64291 Darmstadt, Germany}

\author{Manuel Reichert}
\affiliation{Department of Physics and Astronomy, University of Sussex, Brighton, BN1 9QH, U.K.}

\begin{abstract}
Effective field theories for cosmology offer a powerful framework to investigate the dynamics of space--time and address longstanding open puzzles.
In this work, we initiate a programme to analyse the ultraviolet completion of vector--tensor quantum field theories within the asymptotic safety paradigm, focusing on generalised Proca theories with a vector condensate.
This enables us to assess whether a consistent fundamental UV completion exists and to constrain the set of viable infrared scenarios. Using the non--perturbative functional renormalisation group, we identify several fixed points, including Proca--type candidates, and, among them, a particularly remarkable one with four relevant directions: two associated with gravity and two induced by matter.
This provides evidence for the non--perturbative renormalisability of vector--tensor theories. We further outline how the resulting UV critical surface constrains late--time cosmology.
\end{abstract}

\maketitle

\section{Introduction} 
\label{sec:Introduction} 
Despite the remarkable success of General Relativity (GR) over a wide range of distances and energy scales, from Solar System dynamics to the physics of the early Universe, the combination of high-precision cosmological observations with persistent theoretical puzzles points towards the need for gravity--matter systems beyond GR \cite{Heisenberg:2018vsk}. Such open challenges concern the origin and nature of dark matter and dark energy, the initial singularity, and the incompatibility of GR with quantum mechanics, together with the growing web of so--called cosmological tensions. Taken together, these issues indicate the need for a more fundamental framework in which gravity and the matter sector can be described consistently across all scales. In this context, a question that is particularly worth addressing is: given a cosmologically motivated effective field theory (EFT) for gravity and matter, which regions of its parameter space admit a consistent ultraviolet (UV) completion within quantum field theory (QFT)? Such UV completions within a QFT setup are necessarily asymptotically safe (AS), for recent reviews, see \cite{Bambi:2023jiz, Bonanno:2020bil, Dupuis:2020fhh, Knorr:2022dsx, Eichhorn:2022gku, Morris:2022btf, Wetterich:2022ncl, Martini:2022sll, Saueressig:2023irs, Pawlowski:2023gym, Platania:2023srt, Bonanno:2024xne, Reichert:2020mja, Basile:2024oms}. 

The set of EFTs admitting an interacting UV fixed point (FP) of the coupled gravity--matter system is called the \emph{AS landscape}. In this setting, the renormalisation group (RG) flow describes how the couplings of the effective action evolve with the energy scale, and the presence of such a FP implies that this evolution remains finite and controlled at Planckian and trans--Planckian momentum scales. As a consequence, only a finite number of RG trajectories emanate from the FP towards the infrared (IR), corresponding to a restricted set of low--energy theories and thus to predictive IR scenarios. The complement of the AS landscape is the \emph{AS swampland} and contains the EFTs without UV completion. In this work, we initiate a programme to map out the UV--consistent landscape of cosmological EFTs coupled to gravity.

\begin{figure}[t]
	\centering
	\includegraphics[width=\columnwidth]{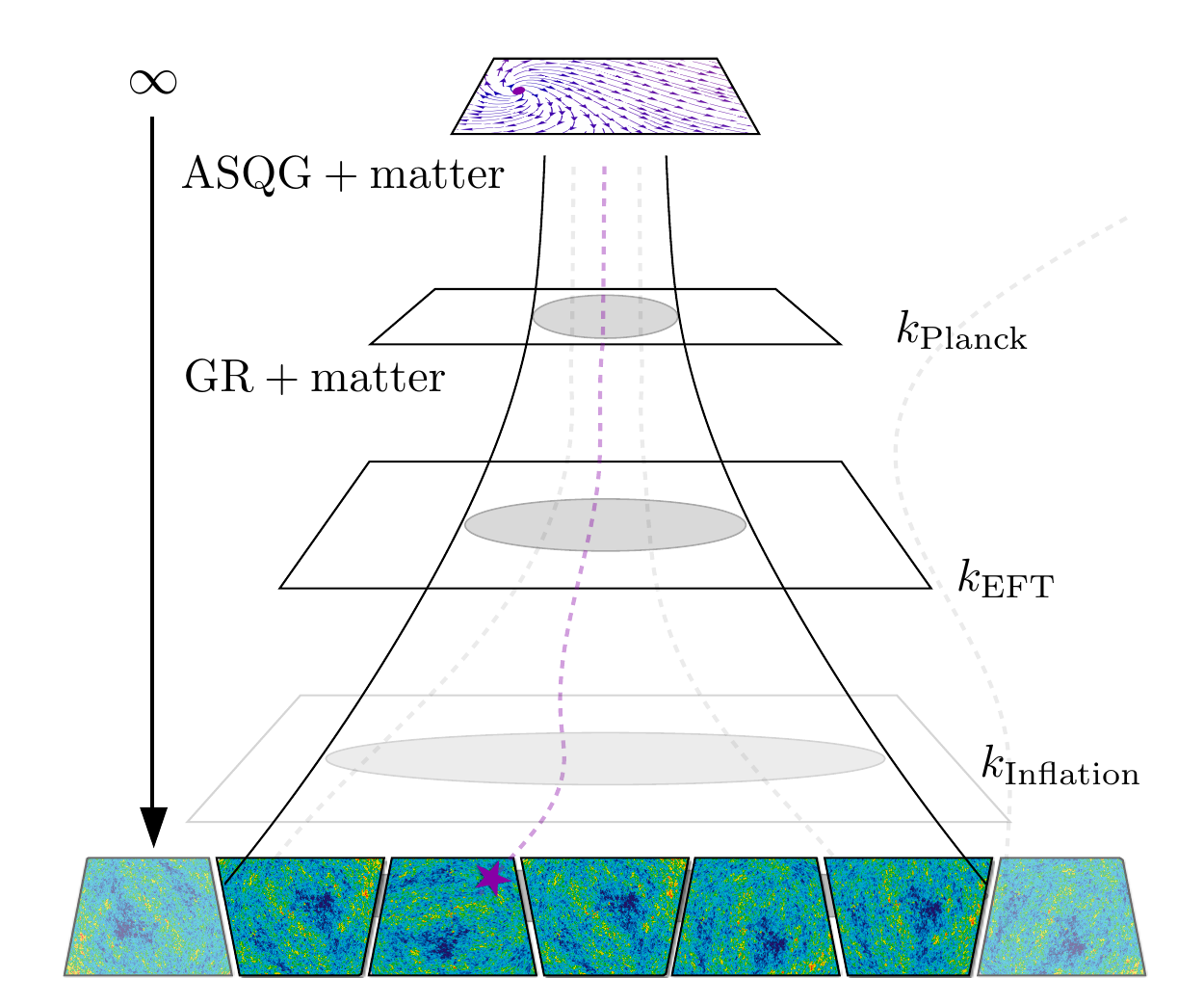}
	\caption{
    Depiction of the embedding of a cosmological EFT within a fundamental UV--complete QFT. The grey areas indicate the landscape of Proca EFTs admitting an AS UV-closure. They are submanifolds of the Proca-EFT surfaces at the Planck scale $k_\textrm{Planck}$ and at some EFT UV scale $k_\textrm{Proca}$. They constrain the physics predictions at $k=0$, indicated with the Cosmic Microwave Background (CMB) background.}
\label{fig:ProcaLandscape+Swamp}
\end{figure}

While a given EFT may be compatible with current cosmological data, it need not admit a consistent extension to asymptotically large momentum and energy scales. Conversely, the subset of UV--complete theories maps onto a restricted set of possible IR observables, thus providing a top--down perspective for disentangling viable extensions of the cosmological Standard Model (SM), in close analogy to the swampland programme \cite{Eichhorn:2024rkc}; an idea we illustrate in \Cref{fig:ProcaLandscape+Swamp}. This programme thus has a twofold purpose: if we only consider a subset of EFTs and we find that the increasing wealth of cosmological data favours a theory in the AS swampland, this would point to the exciting necessity of new physics beyond the EFT set considered. In turn, if the cosmologically compatible EFT lies in the AS landscape, this offers the possibility that late--time cosmology can be described within a minimal extension of the AS Standard Model, see e.g.~\cite{Pastor-Gutierrez:2022nki}.

In this first work, we set up the framework for a cosmologically motivated subset of generalised Proca theories (GPTs) \cite{Heisenberg:2014rta} and perform a FP analysis for the subsequent gravity--Proca system. For that purpose, we consider a massive vector field $A_\mu$ coupled to gravity through a non-minimal truncation including derivative self--interactions that allows for the emergence of a condensate for $A_\mu^2$. Our approximation of the full quantum effective action captures the leading operators that (i) shift the Proca mass, (ii) induce cubic and quartic self--interactions for the vector, and (iii) feed back into the gravitational sector at tree level, while automatically satisfying the stringent gravitational-wave bound on the tensor propagation speed by construction \cite{DeFelice:2016yws, DeFelice:2016uil, deFelice:2017paw, Heisenberg:2016eld, DeFelice:2016cri}.

We identify several Proca--type fixed-point candidates in the AS landscape, featuring different stability properties and connecting to different IR regimes. A particularly interesting fixed-point candidate features four relevant directions: two in the gravity sector and two in the Proca sector. Overall, these FPs provide evidence for the non--perturbative renormalisability of Proca--type EFTs.

\section{Asymptotically safe Proca cosmologies} 
\label{sec:EFTsCosmology}

GPTs \cite{Heisenberg:2014rta} arise as the spin--1 analogue of Horndeski and degenerate scalar--tensor theories, and describe a massive vector with broken $U(1)$ symmetry and derivative self-interactions arranged so that only three healthy polarisations propagate on curved spacetime, with second-order equations of motion and no ghost instabilities, making them viable EFT candidates for describing the dynamics of cosmic acceleration and structure formation \cite{DeFelice:2016uil, DeFelice:2016yws}.

They are used as a cosmological low--energy EFT within an expansion of the theory in local operators ordered by their canonical dimension. The most relevant invariants of the Proca field $A_\mu$ are given by
\begin{align} 
    X&= \frac{Z_A}{2} A_{\mu}A^{\mu}\,, &&\mathrm{and}&
    \Theta&=\frac{Z_A}{4} F_{\mu\nu}F^{\mu\nu}\,,
\label{eq:ProcaInvariantsRel}
\end{align}
where $F_{\mu\nu}=\nabla_{\mu}A_{\nu}-\nabla_{\nu}A_{\mu}$ is the field strength tensor, and $\nabla_{\mu}$ the covariant derivative. The multiplication with powers of the Proca field wave function $Z_A$ renders them RG-invariant. The full quantum effective action for vector--tensor (V--T) theories can be constructed using these invariants and other higher-dimensional ones as building blocks \cite{Heisenberg:2014rta}, and contains an infinite tower of operators. 

In this work, we want to determine the effective action $\Gamma[\bar g_{\mu\nu}, \bar A_\mu,\phi]$ in a gauge--fixed approach to gravity. Accordingly, the metric is split into a flat (Euclidean) background and a graviton fluctuation, namely
\begin{align}
    g_{\mu\nu} = \delta_{\mu\nu} + \sqrt{16\pi G_\text{N}}\ h_{\mu\nu}\,,
\label{eq:MetricExpansion}
\end{align} 
For the Proca sector, we may also introduce a background-fluctuation split for the Proca field, 
\begin{align} 
    A_\mu =\bar A_\mu  + a_\mu\,. 
\label{eq:LinSplitProca}
\end{align}
This leaves us with the quantum effective action
\begin{align} 
    \Gamma[\bar g_{\mu\nu}, \bar A_\mu, \phi] =&\, \int \! \mathrm{d}^4 x\,\sqrt{g}\,\Big\{\Theta+ G_2(X) + G_4(X)\,R \nonumber\\[1ex]\nonumber 
    &\,+ G_{R^2}(X)\,R^2 + G_{C^2}(X)\,C_{\mu\nu\rho\sigma}^2\\[1ex] 
    &\,+ Z^{1/2}_A G_3(X)\,\nabla^\mu A_\mu +\cdots\Big\}\,,
\label{eq:GammaApprox}
\end{align}
where $G_i(X)$ with $i=2,3,4,R^2,C^2$ are general functions of the invariant $X$, and $\phi$ collectively denotes the dynamical fluctuation fields,
\begin{align}
	\phi= (h_{\mu\nu},\, c_\mu,\,\bar c_\mu,\, a_\mu,\,...)\,,
\label{eq:FluctuationFields}
\end{align}
with the metric fluctuation field $h_{\mu\nu} $, the gravity ghosts $\{c_\mu,\,\bar c_\mu\}$ and the Proca field $a_\mu$. The dots stand for further matter and gauge fields that one may consider in a comprehensive theory of high--energy physics and gravity. 

The effective action \labelcref{eq:GammaApprox} includes the effective model used in \cite{DeFelice:2016yws,DeFelice:2016uil} to describe the simplest cosmologically motivated subclass of GPTs. On top of this, it includes additional terms that are generated by quantum fluctuations, such as curvature-square operators, which are not part of the standard generalised Proca construction. Importantly, we treat these terms such that no new (ghost) degrees of freedom are included in the gravitational sector, as detailed in \Cref{sec:FluctuationApproach}. The quantum effective action in \labelcref{eq:GammaApprox} can also include higher-order terms in $\Theta$, which are neglected in this work.

%

\subsection{The UV-landscape of Proca theories}
\label{sec:ScaleDependentProcaAction}

The landscape of AS gravity--Proca EFTs is determined through a UV FP analysis. The number of relevant directions provides us with the number of coupling parameters of these theories, and by scanning their corresponding ranges and allowed values, we can map out the maximal set of UV--complete EFTs, as depicted in \Cref{fig:ProcaLandscape+Swamp}. Such an analysis is best done within the functional renormalisation group (fRG) approach to AS gravity--matter systems, in which one studies the scale dependence of an IR-regulated effective action $\Gamma_k$ that interpolates between a microscopic action in the UV and the full quantum effective action $\Gamma=\Gamma_{k\to 0}$ in the IR; for a comprehensive collection of recent reviews see \cite{Bambi:2023jiz, Bonanno:2020bil, Dupuis:2020fhh, Knorr:2022dsx, Eichhorn:2022gku, Morris:2022btf, Wetterich:2022ncl, Martini:2022sll, Saueressig:2023irs, Pawlowski:2023gym, Platania:2023srt, Bonanno:2024xne, Reichert:2020mja, Basile:2024oms}. 
In this approach, the RG flow of the scale--dependent effective action $\Gamma_k$, or that of an approximation such as \labelcref{eq:GammaApprox}, is governed by the flow equation \cite{Wetterich:1992yh},
\begin{subequations}
\label{eq:FunFlow}
\begin{align}
	\partial_t \Gamma_k[\bar g_{\mu\nu}, \bar A_\mu,\phi]= \frac{1}{2}\,{\rm Tr} \! \left[\frac{1}{ \Gamma_k^{(2)}[\bar g_{\mu\nu}, \bar A_\mu,\phi]+R_k}\partial_t R_k \right]\,,
\label{eq:FlowEffAct}
\end{align}
where $t$ denotes the (negative) RG-time, $t=\ln{(k/k_{\rm ref})}$, with $k_\textrm{ref}$ a fixed reference scale. The right-hand side involves the regulated, connected part of the full two-point function, $1/(\Gamma_k^{(2)}+R_k)$, and depends on the one-particle irreducible vertices $\Gamma^{(n)}$ obtained from functional derivatives of the effective action
\begin{align}
    \Gamma^{(n)}_{\phi_{i_1}\cdots\phi_{i_n}}[\bar g_{\mu\nu}, \bar A_\mu,\phi](p_1,\ldots,p_n)= \frac{\delta^n \Gamma[\bar g_{\mu\nu}, \bar A_\mu,\phi]}{\delta \phi_{i_1}(p_1)\cdots \delta \phi_{i_n}(p_n)}\,.
\label{eq:nPointfunction}
\end{align}
\end{subequations}
The remaining ingredient in \labelcref{eq:FunFlow} is the regulator function $R_k(p)$, which acts as a momentum--dependent mass term that suppresses quantum fluctuations below a given IR cutoff scale $k$. It is chosen such that it vanishes in the UV, $p^2/k^2\to\infty$, while it remains finite---or even diverges---in the IR, $p^2/k^2\to 0$. In this way, the transition between these two regimes takes place around $p^2/k^2\simeq 1$, and the full quantum effective action $\Gamma[\phi]=\Gamma_{k\to 0}[\phi]$ is obtained by successively integrating out fluctuations as the IR cutoff scale $k$ is lowered.

For the derivation of the RG flows and a FP analysis, it is convenient to write the effective action in terms of the dimensionless counterparts $g_i$ of the couplings $G_i$,
\begin{align} 
    g_i &= k^{-d_i} G_i\,, 
    &
    d_i &= [G_i]\,,
\label{eq:DimlessGs}
\end{align} 
where $d_i$ denotes the canonical mass dimension of $G_i$. The scale--dependent effective action can be viewed as a functional of the fields parametrised by the coupling vector $\boldsymbol{g}$,
\begin{align} 
    \Gamma[\bar g_{\mu\nu}, \bar A_\mu, \phi; \boldsymbol{g}] = 
    \Gamma_k[\bar g_{\mu\nu}, \bar A_\mu, \phi]\,,\quad  
    \boldsymbol{g}=(g_1,\,g_2,\,...)\,,
\label{eq:Gammag}
\end{align} 
where fixing the RG scale $k$ is equivalent to fixing the values of the dimensionless couplings $\boldsymbol{g}$. In this form, the scale dependence of the theory is fully encoded in the running of the set $\boldsymbol{g}$.

FPs are given by the zeros of the flow of the effective action, and imply its quantum scale invariance. This appears as the freezing of the dimensionless couplings $g_i$, while the dimensionful versions $G_i$ scale according to their classical dimension. FP solutions $\boldsymbol{g}^*=(g_1^*,\,g_2^*,\,\ldots)$ are obtained as the roots of the complete set of $\beta$-functions
\begin{align}
    \boldsymbol{\beta}(\boldsymbol{g}^*) = 0\,, \qquad \textrm{with} \qquad
    \beta_{g_i}(\boldsymbol{g}) = \partial_t g_i\,.
\label{eq:FP}
\end{align}
Here, $\boldsymbol{\beta}=(\beta_{g_1},\beta_{g_2},\ldots)$ denotes the set of beta functions of the dimensionless couplings. They can be written as
\begin{align}
    \beta_{g_i} = - d_i\, g_i + \mathrm{Flow}_i(\boldsymbol{g})\,,
\label{eq:beta_def}
\end{align}
where the first term encodes the canonical scaling and $\mathrm{Flow}_i(\boldsymbol{g})$ contains the quantum contributions obtained from the right-hand side of \labelcref{eq:FlowEffAct}. The latter include the anomalous dimensions
\begin{align}
    \eta_i := - \partial_t \ln Z_i \,,
    \label{eq:AnomalousDimDef}
\end{align}
associated with the corresponding wave--function renormalisations $Z_i$, as introduced in \Cref{sec:EFTsCosmology}, and the quantum corrections to each of the respective vertices. 

\begin{figure}[t]
	\centering
	\includegraphics[width=\columnwidth,
    clip, trim=45pt 75pt 30pt 100pt 
    ]{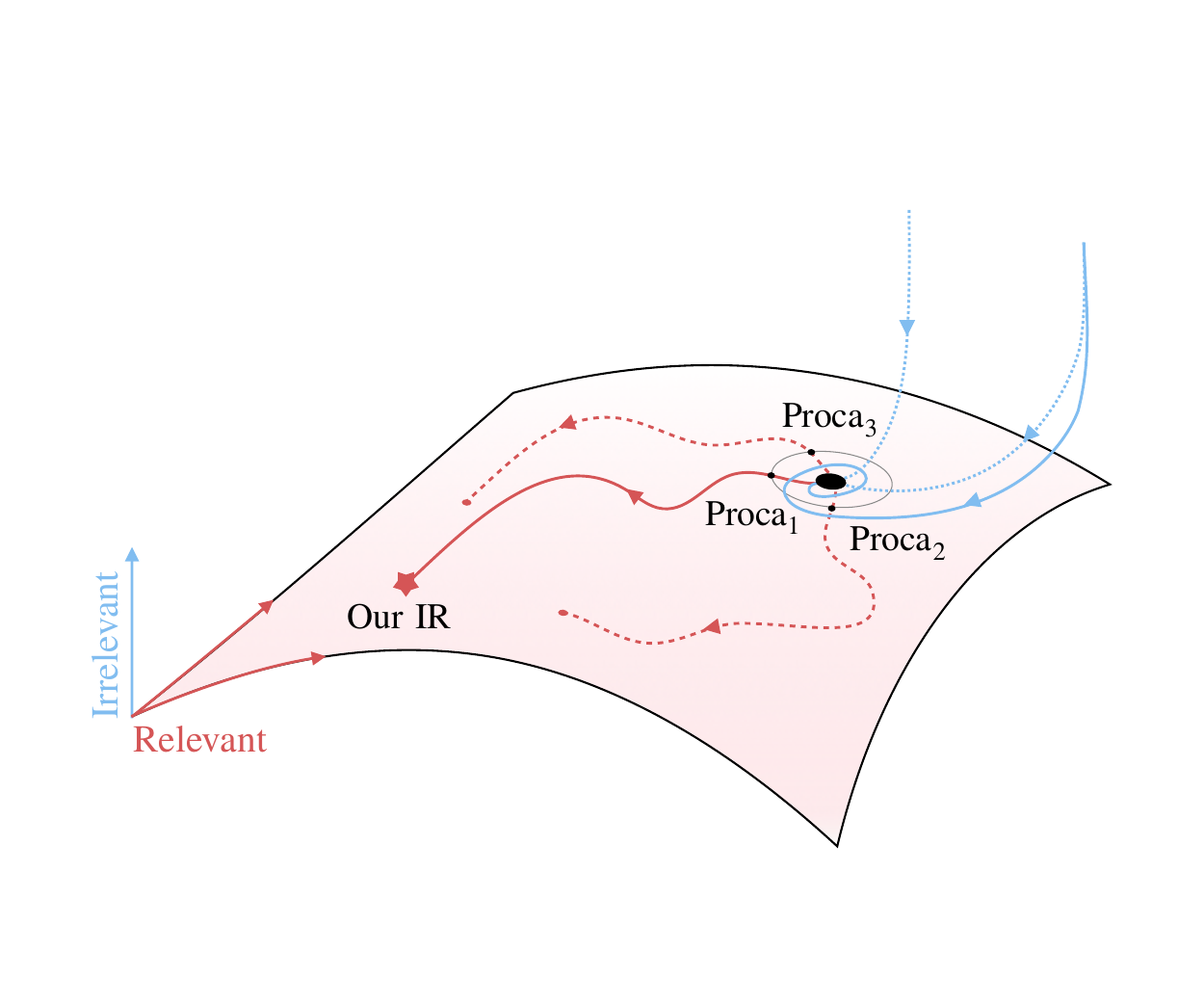}
    \caption{Schematic depiction of predictivity in AS theories. An interacting UV FP controls the high--energy behaviour of the theory. Relevant directions (red) span the UV--critical surface and correspond to couplings that must be fixed by IR data, while irrelevant directions (blue) are attracted towards the FP and are therefore predicted. Renormalisation group trajectories flowing away from the FP define different EFTs at low energies. Proca$_{1,2,3}$ denote representative RG trajectories emanating from the same UV FP along different choices of relevant couplings, yielding distinct IR Proca EFT realisations.}
\label{fig:ProcaPredictivity}
\end{figure}
%

\subsection{Predictivity of Proca theories}
\label{sec:PredictivityProca}

Predictive theories have UV FPs \labelcref{eq:FP} with a finite number of relevant and marginal directions (UV-attractive) and an infinite number of irrelevant directions (UV-repulsive). This information can be extracted from the stability matrix $B$, obtained by linearising the $\beta$-functions around the FP $\boldsymbol{g}^*$. The matrix elements and the eigenvalues of the stability matrix are given, in our convention, by 
\begin{align}
    B_{ij} = \left.\frac{\partial \beta_{g_i}}{\partial g_j}\right|_{g=g_*},\qquad \theta_I =\,\text{eig}(B)\,. 
\label{eq:StabilityMatrix}
\end{align}    
The eigenvalues or critical exponents $\theta_I$ classify the relevant and irrelevant directions: Eigenvalues with $\text{Re}\,\theta_I<0$ correspond to relevant operators and the respective couplings must be fixed to render the desired IR. Those with $\text{Re}\,\theta_I>0$ correspond to irrelevant operators and their evolution is fully determined by the relevant operators. A predictive UV completion is given when the number of relevant directions is finite and implies that the theory is (nonperturbatively) renormalisable. 

The critical exponents are universal quantities that dictate the behaviour of small deformations of the theory in the vicinity of a FP. A relevant direction corresponds to a perturbation that grows as the RG scale is lowered towards the IR: its associated coupling must therefore be fixed by experimental or observational input at low energies. By contrast, an irrelevant direction is attracted back towards the FP as the flow proceeds to the UV, and the corresponding coupling is dynamically predicted once the relevant parameters are specified. In this sense, the FP acts as an organising centre in theory space: the UV--critical surface is spanned by the relevant directions, while irrelevant directions are fixed functions of them. Consequently, the smaller the number of relevant directions, the higher the predictive power of the theory, as fewer independent parameters need to be measured in the IR. This structure is illustrated schematically in \Cref{fig:ProcaPredictivity}, where RG trajectories emanating from the UV FP flow into distinct EFTs at low energies. For a comprehensive discussion, we again refer to the reviews \cite{Bambi:2023jiz, Bonanno:2020bil, Dupuis:2020fhh, Knorr:2022dsx, Eichhorn:2022gku, Morris:2022btf, Wetterich:2022ncl, Martini:2022sll, Saueressig:2023irs, Pawlowski:2023gym, Platania:2023srt, Bonanno:2024xne, Reichert:2020mja, Basile:2024oms}.

The UV--closure and predictivity of the Proca--gravity theories are also illustrated in \Cref{fig:ProcaLandscape+Swamp}. 
At a given scale $k$, the landscape of Proca EFTs with an AS UV--closure is the set ${\cal L}_k$ of couplings $\boldsymbol{g}_k$ for a given effective action $\Gamma[\bar g_{\mu\nu}, \bar A_\mu,\phi;\boldsymbol{g}_k]$. 
Assume now that we have found a Proca EFT with an effective action $\Gamma_{k_\textrm{Proca}}$ at a finite UV scale $k_\textrm{Proca}$ that describes the experimental data. This theory is defined by couplings $\boldsymbol{g}_{k_\textrm{Proca}}$. 
\\[-1ex]

\textit{(i)}: If $\boldsymbol{g}_{k_\textrm{Proca}}$ lies in the AS Proca landscape ${\cal L}_{k_\textrm{Proca}}$ at $k_\textrm{Proca}$, 
\begin{align} 
    \boldsymbol{g}_{k_\textrm{Proca}}\in{\cal L}_{k_\textrm{Proca}}\,,
\label{eq:ASsafeProca}
\end{align} 
the respective Proca EFT has an AS UV--closure. \\[-1ex] 

\textit{(ii)}: If $\boldsymbol{g}_{k_\textrm{Proca}}$ lies in the complement of ${\cal L}_{k_\textrm{Proca}}$, %
\begin{align} 
    \boldsymbol{g}_{k_\textrm{Proca}}\notin{\cal L}_{k_\textrm{Proca}}\,,
\label{eq:ASunsafeProca}
\end{align} 
the respective Proca EFT is asymptotically unsafe. \\[-1ex]

Both outcomes are exciting: \textit{(i)} hints at a minimal UV--closure of particle physics and cosmology in terms of an AS GPT. In turn, \textit{(ii)} points towards new physics beyond the cosmological SM. 

Moreover, in an AS setting, predictivity is quantified by the codimension of the UV--critical surface. 
A small number of relevant parameters implies correlations among late--time observables (e.g.\ background expansion and linear--response coefficients). 
From the cosmological perspective, the programme is therefore twofold: (1) isolate the FPs and their relevant directions in the combined gravity--matter theory space; and (2) identify those trajectories that remain compatible with stability and causality conditions on cosmological backgrounds and with observational bounds after RG evolution to the IR. 
From this viewpoint, UV consistency can translate into correlations among EFT parameters and thus reduce the available cosmological parameter space compared to a purely bottom-up EFT analysis.

\section{Fluctuation approach and truncation setup}
\label{sec:FluctuationApproach}

In the present work, we evaluate GPTs in a low-order approximation of the full effective action in \labelcref{eq:GammaApprox}. We only consider canonically relevant and marginal terms, and drop higher orders in the difference $X-X_0$ of the $A_\mu A^\mu$ invariant and its expectation value 
\begin{align}
X_0=\langle X\rangle\,.
\label{eq:X-X0}
\end{align}
The $X$-dependent couplings $G_i(X)$ introduced in \labelcref{eq:GammaApprox} are Taylor expanded in $X-X_0$, 
\begin{align} 
    G_2(X) = &\, G_{20} + M_A^2 X + G_{22}(X-X_0)^2\,, \notag \\[1ex] 
    G_3(X) = &\,G_{31}(X-X_0)\,,\notag\\[1ex]  
    G_4(X) = &\, G_{40}+ G_{41}(X-X_0)\,,\notag\\[1ex] 
    G_{R^2}(X)=&\,G_{R^20} \,,\notag\\[1ex] 
    G_{C^2}(X)=&\,G_{C^20} \,.
\label{eq:GFunExpansion}
\end{align}
\Cref{eq:GammaApprox,eq:GFunExpansion} accommodate all canonically UV-relevant and marginal operators in a GPT. We note that we have relaxed the Proca constraints between the different terms in the effective action in the spirit of the programme of scanning the landscape of UV--complete EFTs. The Proca mass term is introduced in a separate way in $G_2(X)$, while the rest of it acts as an effective potential expanded around $X_0$. By construction, the expansion contains no linear term, ensuring that $X_0$ corresponds to an extremum of the potential. Functional derivatives of such a potential with respect to $A_\mu$ yield contributions proportional to powers of the field, which vanish upon evaluation at the expansion point \labelcref{eq:X-X0} and fluctuation background $\bar A_\mu=0$. Consequently, the potential does not contribute to the vector mass, which is entirely encoded in the explicit mass term singled out in $G_2$, as shall be further clarified in \Cref{sec:FluctuationApproach}.

In this first analysis, we drop the marginal operators $R^2$, $C^2$ and $A^2R$, with their respective couplings $G_{R^2 0}$, $G_{C^2 0}$, and $G_{41}$. The curvature-square operators are not part of the standard generalised Proca construction. More precisely, we treat the curvature-squared operators such that within the graviton propagator they are fully captured by the wave functions $Z_h$ as is standard in the fluctuation approach to quantum gravity \cite{Pawlowski:2020qer, Pawlowski:2023gym}, and neglect all vertex contributions from these operators. This ensures that no new on-shell degrees of freedom are introduced, such as a ghost from the $C^2$-operator or a Starobinsky-type massive scalar mode from the $R^2$-operator. The $A^2R$-operator is also omitted, by analogy to the simplest cosmologically motivated subclass of GPTs, in which $G_4$ is constant. These operators will be included in forthcoming work. 

In summary, the truncation \labelcref{eq:GammaApprox} with \labelcref{eq:GFunExpansion} and $G_{R^2 0}=G_{C^2 0}=G_{41}=0$ is sufficiently rich to capture relevant V--T dynamics, while remaining compatible with standard cosmological viability conditions, including the absence of ghosts and the luminal propagation of gravitational waves by construction. The FP analysis of the effective action is done within the fRG approach introduced in \Cref{sec:ScaleDependentProcaAction}. In this approach, all couplings $G_{ij}$ in \labelcref{eq:GammaApprox,eq:GFunExpansion} depend on the renormalisation scale $k$. This RG scale can be understood as an average momentum scale and agrees well with the symmetric--point momentum of the respective vertices. For detailed discussions of this relation, see in particular \cite{Denz:2016qks} and the reviews \cite{Pawlowski:2020qer, Pawlowski:2023gym}.

\Cref{eq:GammaApprox} with \labelcref{eq:GFunExpansion} includes a pure quantum gravity part with scale--dependent cosmological constant and Newton coupling,  
\begin{align} 
    G_{20} &= \frac{2 \Lambda}{16\pi\, G_\text{N}} \,,
    &
    G_{40} &= -\frac{1}{16\pi\, G_\text{N}}\,.
    \label{eq:Gnm}
\end{align} 
Moreover, \labelcref{eq:GammaApprox} already allows us to accommodate the cosmological dynamics qualitatively: an isotropic FLRW background has been shown to arise naturally for a purely timelike condensate of the Proca field, or more precisely that of the invariant $X$ \cite{DeFelice:2016yws,DeFelice:2016uil,deFelice:2017paw,Heisenberg:2016eld}. In the present work we treat $X$ similarly to a condensate field that accommodates \labelcref{eq:X-X0}. Within such a setup, the expectation value of $A_\mu$ vanishes, which is interpreted as a fluctuation about this background. A fully self-consistent condensate approach also necessitates kinetic terms for the condensate field $X$ which we neglect here. 

In the class of cosmological solutions of \labelcref{eq:GFunExpansion,eq:GammaApprox}, one can formulate simple, mode-by-mode viability criteria for tensor, vector, and scalar perturbations throughout cosmic history. These criteria translate into restrictions on combinations of $G_2(X)$, $G_3(X)$, and $G_4(X)$ and their $X$-derivatives at $X_0$, or rather the set $\boldsymbol{G}$ of scale--dependent couplings $G_{ij}$ with
\begin{align} 
    \boldsymbol{G} = \left(X_0,\,M_a^2,\,G_{20},\,G_{22},\,G_{31},\, G_{40},\,...\right)\,,
\label{eq:g-Proca} 
\end{align} 
which delineate the parameter space compatible with late--time acceleration and structure growth \cite{DeFelice:2016uil,DeFelice:2016yws}. 

For the computation of the $\beta$-functions, we use the vertex expansion scheme that underlies the fluctuation approach to AS gravity--matter systems, see \cite{Pawlowski:2020qer, Pawlowski:2023gym} for reviews. In this framework, the central building blocks are the $n$--point vertices $\Gamma^{(n)}$, evaluated on a chosen background metric. 

Concerning the vertex expansion in the gravity sector, we use the Einstein--Hilbert tensor structures dressed with a single wave function $Z_h$ of the transverse-traceless fluctuation graviton. While the tensor structures are generated from the Einstein--Hilbert action, the momentum dependence of the wave function encodes overlap with higher-order operators. On the level of the propagator, this implies that we use the full tensor basis and we retain a single graviton mass parameter $\mu_h$ computed from the transverse-traceless mode. For higher $n$-point functions, we have the dimensionless couplings $g_{h^n}$ and $\lambda_{h^n}$ which are avatars of the Newton coupling and cosmological constant. We have further avatars of the Newton coupling associated with mixed graviton-Proca vertices $g_{h^n a^m}$. Following standard practice (see \cite{Pawlowski:2020qer, Pawlowski:2023gym}), we identify all gravitational couplings with those of the three-graviton vertex,
\begin{align}
g_{h^n} &= g_{h^n a^m} = g_{h^3},
&
\lambda_{h^n}&=\lambda_{h^3}; 
&
&\mathrm{for}\; n>3.
\end{align} 
This identification is inspired by classical diffeomorphism invariance and justified on the quantum level by effective universality \cite{Eichhorn:2018akn, Eichhorn:2018ydy}.

\begin{table*}[t]
    \renewcommand{\arraystretch}{1.1}
\begin{ruledtabular}
\begin{tabular}{l c c c c c c c l}
        FP & $\xi$ & $g_{h^3}$ & $\mu_h$ & $\lambda_{h^3}$ & $\mu_a$ & $g_{a^3}$ & $g_{a^4}$ & $\{\theta_i\}$ \\%
        \midrule
        \midrule
        \multirow{4}{*}{\shortstack[l]{Minimally\\coupled}}
        & 0 & 0.54 & $-0.15$ & $-0.19$ & 0 & 0 & 0 & $\{-1.5\pm1.6\,i,\,1.4;\,-2.1,\,-0.48,\,-0.17\}$ \\
        & 1 & 0.54 & $-0.15$ & $-0.19$& 0 & 0 & 0 & $\{-1.5\pm1.6\,i,\,1.4;\,-2.1,\,-0.48,\,-0.17\}$ \\
        & 100 & 0.54 & $-0.15$ & $-0.19$& 0 & 0 & 0 & $\{-1.5\pm1.6\,i,\,1.4;\,-2.1,\,-0.48,\,-0.17\}$ \\
        & $\infty$ & 0.54 & $-0.15$ & $-0.19$ & 0 & 0 & 0 & $\{-1.5\pm1.6\,i,\,1.4;\,-2.1,\,-0.48,\,-0.17\}$ \\
        \midrule
        \multirow{4}{*}{Interacting} 
        & 0 & 0.53 & $-0.15$ & $-0.19$ & $-0.12$ & 0 & $7.6$ & $\{-1.5\pm1.6\,i,\,1.4;\,-1.3,\,-0.35,\,0.43\}$ \\
        & 1 & 0.54 & $-0.15$ & $-0.19$ & $-0.094$ & 0 & $4.6$ & $\{-1.5\pm1.6\,i,\,1.5;\,-1.4,\,-0.37,\,0.31\}$ \\
        & 100  & 0.54 & $-0.15$ & $-0.19$ & $-2.4\cdot 10^{-4}$ & 0 & $4.9\cdot 10^{-4}$ & $\{-1.5\pm1.6\,i,\,1.5;\,-2.0,\,-0.32,\,0.18\}$ \\
        & $\infty$  & 0.54 & $-0.15$ & $-0.19$ & 0 & 0 & 0 & $\{-1.5\pm1.6\,i,\,1.5;\,-2.0,\,-0.31,\,0.18\}$ \\
        \midrule
        \multirow{3}{*}{Gemini$_1$} 
        & 10 & $0.51$ & $-0.22$ & $-0.14$ & $-0.19$ & $\pm\, 1.1$ & $0.87$ & $\{-2.1\pm2.2\,i,\,1.9;-5.0,\,6.1\pm7.4\,i\}$\\
        & 100 & $0.53$ & $-0.20$ & $-0.16$ & $-0.020$ & $\pm\, 0.16$ & $0.018$ & $\{-1.8\pm1.9\,i,\,1.8;-6.3,\,5.7\pm6.8\,i\}$\\
        & $\infty$ & $0.53$ & $-0.20$ & $-0.16$ & 0 & 0 & 0 & $\{-1.8\pm1.9\,i,\,1.8;-6.5,\,5.7\pm6.8\,i\}$\\
        \midrule
        \multirow{2}{*}{Gemini$_2$} 
        & 100 & $0.54$ & $-0.14$ & $-0.19$ & $0.0079$ & $\pm\, 0.26$ & $0.15$ & $\{-1.5\pm1.6\,i,\,1.4;\,-1.2\pm1.1\,i,\,-15\}$\\
        & $\infty$ & $0.54$ & $-0.14$ & $-0.19$ & 0 & 0 & 0 & $\{-1.5\pm1.6\,i,\,1.4;\,-1.1\pm1.1\,i,\,-11\}$\\
        \midrule
        \multirow{3}{*}{\shortstack[l]{Proca$^\star$}}
        & 10 & 0.60 & $-0.29$ & $-0.09$ & $-0.51$ & 0 & 21 & $\{-1.0\pm2.8\,i,\,2.4;\,-7.3,\,-4.2,\,17\}$ \\
        & 100 & $0.54$ & $-0.20$ & $-0.16$ & $-0.37$ & 0 & 17 & $\{-1.2\pm1.7\,i,\,1.6;\,-2.0,\,-1.0,\,5.3\}$ \\
        & $\infty$ & ${0.53}$ & ${-0.18}$ & ${-0.17}$ & ${-0.32}$ & ${0}$ & ${15}$ & ${\{-1.3\pm1.7\,i,\,1.5;\,-1.3,\,-0.70,\,3.6\}}$ \\

\end{tabular}
\end{ruledtabular}
\caption{Summary of the fixed-point coordinates, $\boldsymbol{g}^*$, and critical exponents, $\theta_i$, for the FPs of the gravity--Proca system, listed for representative values of the Proca pseudo--gauge fixing parameter $\xi$.}
\label{tab:fixed_points_summary}
\end{table*}

In the Proca sector, we include the Proca wave--function renormalisation $Z_a$ and the Proca mass, expressed in terms of the dimensionless mass parameter $\mu_a:=M_a^2/k^2$; as well as the cubic and quartic Proca self-interactions, parametrised by the couplings $g_{a^3}$ and $g_{a^4}$. Furthermore, we assume the irrelevance of higher $n$--point functions, 
\begin{align}
    g_{a^n}=0, \qquad \mathrm{for}\; n\geq 5.
    \label{eq:ProcaCouplingsIdentif}
\end{align} 
The final ingredient of our truncation is a pseudo--gauge fixing of the Proca propagator. This extension allows us, in the first place, to also accommodate, within the same framework, UV FPs that resemble those of Abelian and non-Abelian gauge theories. In addition, the parameter $\xi$ acts as a regulator and controls the longitudinal sector of the vector propagator, allowing us to monitor the role of longitudinal fluctuations in the flow (see \Cref{app:ProjectedFlowEquations} for details on the regularisation). The corresponding generic Proca two-point function is then parametrised as
\begin{align} 
    \left[\Gamma^{(2)}_{aa}\right]_{\mu\nu} = Z_a(p)\,\delta_{\mu\nu}\left(p^2 + M_a^2\right) + Z_a^\parallel(p)\left(1-\frac{1}{\xi}\right)p_\mu p_\nu \,,
\label{eq:Procaaa} 
\end{align}
and the Proca theory is recovered in the limit $\xi\to\infty$. This approach has further been used in the study of vector mesons in QCD \cite{Jung:2019nnr}. In contrast, the Landau gauge $\xi=0$ is usually used to study the FPs of AS gauge theories since $\xi=0$ is itself a FP of the flow, see \cite{Litim:1998qi}. This is discussed in more detail in \Cref{sec:Results}. In contrast, the limit $1/\xi\to 0$ is not protected by symmetries. In the present work, we therefore treat $\xi$ as an external parameter, and we perform a FP scan as a function of $\xi$ in order to explore how interacting FPs deform between these limits. Moreover, for simplicity, we set $Z_a^\parallel(p)=Z_a(p)$ and thus describe both polarisation sectors with a single wave--function renormalisation. 

Overall, we compute the flow of the graviton two- and three-point functions $\Gamma^{(2h)}$, $\Gamma^{(3h)}$, and the flow of the Proca two-, three-, and four-point functions $\Gamma^{(2a)}$, $\Gamma^{(3a)}$, $\Gamma^{(4a)}$. Technical details on the conventions and projection schemes used for the computation of the flows are provided in \Cref{app:ProjectedFlowEquations}. Altogether, this setup leaves us with the following set of dimensionless running couplings,
\begin{align}
    \boldsymbol{g}
    = \left(\mu_h,\,g_{h^3},\,\lambda_{h^3},\,\mu_a,\,g_{a^3},\,g_{a^4},\,\ldots\right)\,.
\label{eq:boldg-FluctuationApproach}
\end{align}
These are obtained from the truncation \labelcref{eq:GammaApprox} via the expansion \labelcref{eq:GFunExpansion}, and within the Proca sector, are in a one-to-one correspondence with the set of couplings $\boldsymbol{G}$ in \labelcref{eq:g-Proca} via the map:
\begin{align}
    M_a^2  =&\,\mu_a k^2\,,
    &
    G_{31}  =&\,2\,g_{a^3}\,,
    &
    G_{22} =&\, g_{a^4}\,.
    \label{eq:Gtog}
\end{align}
The Proca condensate $X_0$ is fixed by requiring that the effective potential for the invariant $X$ admits an extremum at $X=X_0$. For $X_0\neq 0$, this condition yields
\begin{align}
    X_0 = -\frac{\mu_a}{2\,g_{a^4}}\,k^2.
\label{eq:x0_condition}
\end{align}
As expected, the dependence on $X_0$ is absorbed into the fluctuation couplings and does not appear explicitly in the flow equations.

Within the gravity sector, we do not have a one-to-one correspondence since we have two avatars of the cosmological constant, the graviton mass parameter $\mu_h$ and the constant part of the three-graviton vertex $\lambda_{h^3}$.

\begin{figure*}[t]
\centering
\includegraphics[width=0.75\textwidth]
{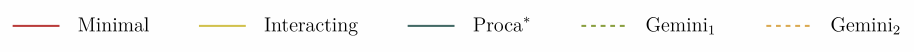}
\makebox[\textwidth]{%
\begin{minipage}{0.49\textwidth}
    \centering
    \includegraphics[width=\linewidth]{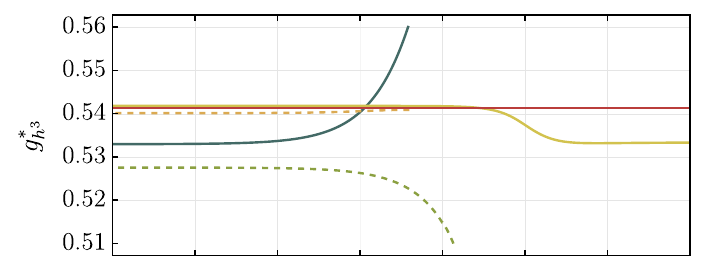}
    \includegraphics[width=\linewidth]{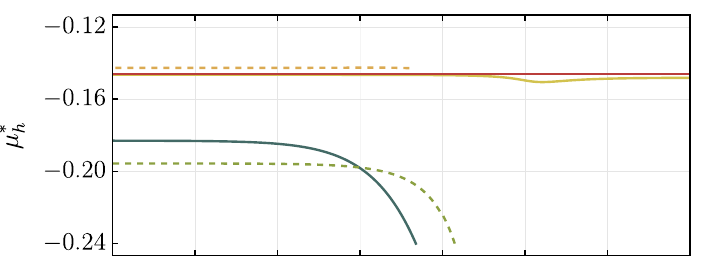}
    \includegraphics[width=\linewidth]{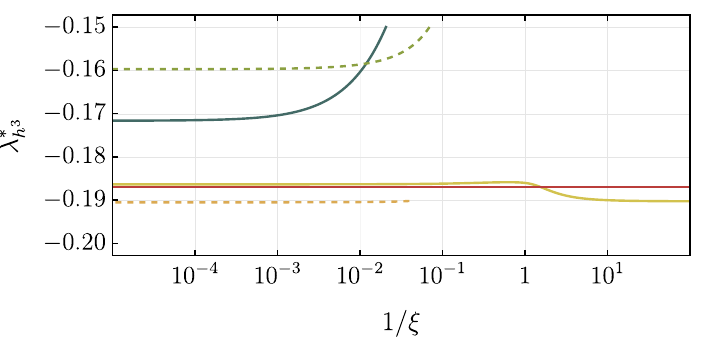}
\end{minipage}\hfill
\begin{minipage}{0.49\textwidth}
    \centering
    \includegraphics[width=\linewidth]{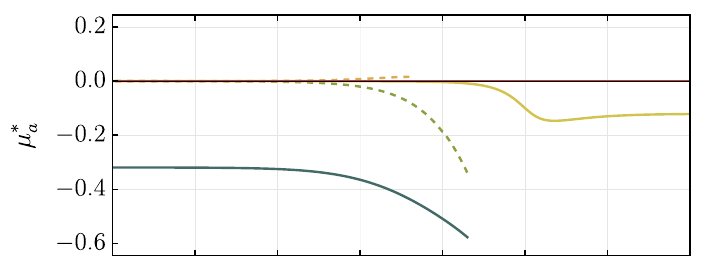}
    \includegraphics[width=\linewidth]{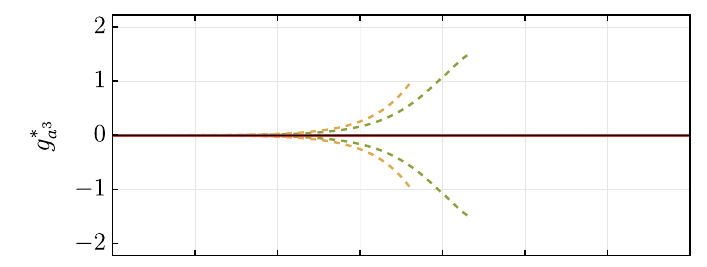}
    \includegraphics[width=\linewidth]{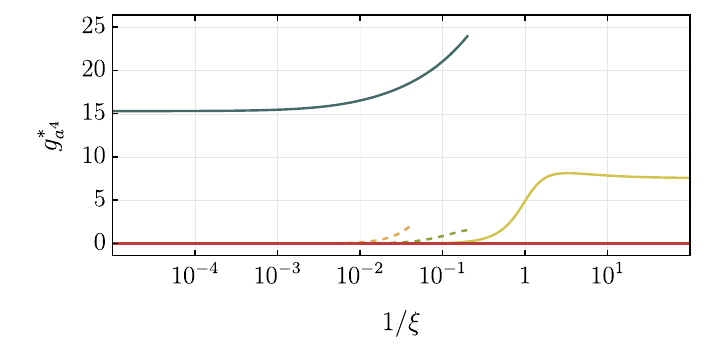}
\end{minipage}
}
\caption{Fixed-point values of the gravity couplings $g_{h^3}^*$, $\mu_h^*$, and $\lambda_{h^3}^*$ (left), and the Proca couplings $\mu_a^*$, $g_{a^3}^*$, and $g_{a^4}^*$ (right), as functions of $1/\xi$ (log scale) for the minimally coupled, interacting, interacting Proca$^\star$, Gemini$_1$ (values plotted for the regime $\xi \geq 5$) and Gemini$_2$ (values plotted for the regime $\xi \geq 25$) FPs.}
\label{fig:ProcaFPscan_all_withGemini}
\end{figure*}
\begin{figure*}[t]
	\centering
	\makebox[\textwidth]{%
\begin{minipage}{0.49\textwidth}
    \centering
    \includegraphics[width=\columnwidth]{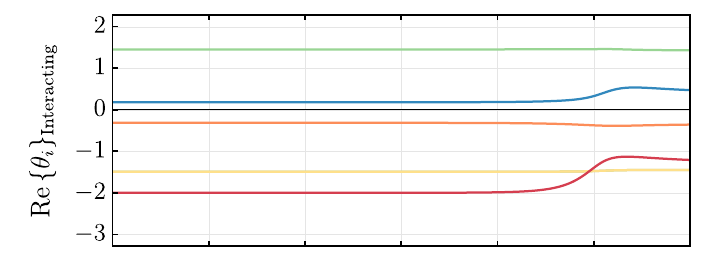}
    \includegraphics[width=\columnwidth]{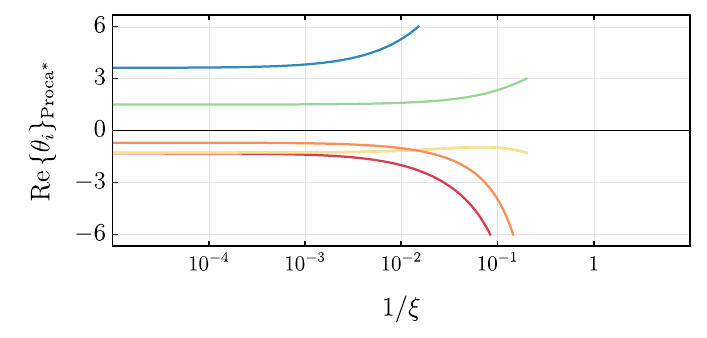}
\end{minipage}\hfill
\begin{minipage}{0.49\textwidth}
    \centering
    \includegraphics[width=\columnwidth]{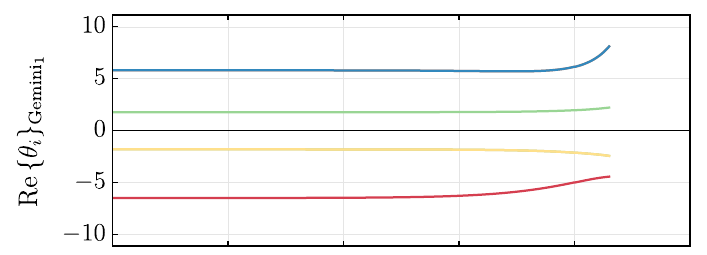}
	\includegraphics[width=\columnwidth]{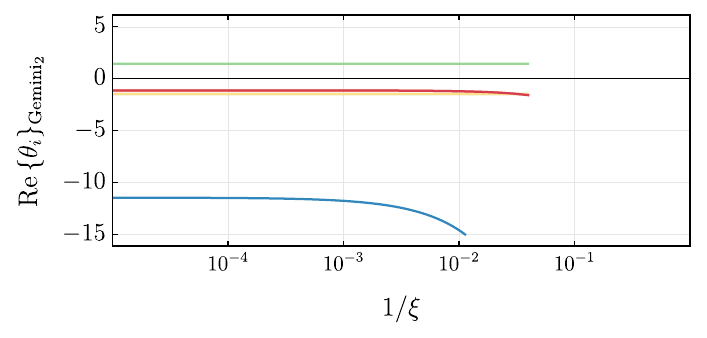}
\end{minipage}
}
\caption{Real parts of the critical exponents, $\mathrm{Re}\{\theta_i\}$, as functions of $1/\xi$ (log scale) for the FPs interacting (top-left), Proca$^\star$ (bottom-left), Gemini$_1$ (top-right, values plotted for the regime $\xi \geq 5$) and Gemini$_2$ (bottom-right, values plotted for the regime $\xi \geq 25$). Lines of the same colour correspond to critical exponents associated with the same eigendirection.}    
\label{fig:FPs_CritExp}
\end{figure*}
%

\section{Fixed points}
\label{sec:Results}

In this section, we discuss the fixed-point solutions for the approximation with the set of couplings $\boldsymbol{g}$ in \labelcref{eq:boldg-FluctuationApproach}. In total, we find five FPs with different physics properties, see \Cref{tab:fixed_points_summary}. The different solutions can be organised into two classes according to their domain of existence in the pseudo--gauge fixing parameter $\xi$. The first class contains FPs that exist for the full range $\xi\in\mathbb{R}^+$, namely the \textit{minimally coupled} and \textit{interacting} Proca--type branches. The second class contains FPs that exist in the strict Proca limit $\xi \to \infty$ but only above a minimum value $\xi>\xi_{\textrm{min}}^{\textrm{FP}_i}$. This class includes two Gemini branches, \textit{Gemini$_1$} with $\xi^{\text{Gemini}_1}_{\rm min}\approx 5$ and \textit{Gemini$_2$} with $\xi^{\text{Gemini}_2}_{\rm min}\approx 25$. Finally, there is an additional remarkable solution that exists for $\xi > \xi_{\rm min}^{\text{Proca}^{\star}}\approx 4$, which, given its characteristics, we shall label as \textit{interacting Proca$^{\star}$}. 
While potential Proca--type theories are obtained in the limit $\xi\to\infty$, the FPs are potentially interesting for other values of $\xi$ in view of the programme of mapping out asymptotically safe EFTs, as discussed in the introduction. In the following \Cref{sec:MinimallyCoupled,sec:Interacting,sec:Gemini,sec:Isolated} we briefly discuss each of the different FPs.  

Our work is complementary to \cite{Heisenberg:2026rdk}, where the same class of theories has been addressed. We note that our study is performed in the fluctuation approach, while \cite{Heisenberg:2026rdk} employs the background-field approximation. The latter has been assessed and embedded in the fluctuation approach, and for minimally coupled gravity--matter systems the ensuing approximation artifacts have been studied in \cite{Meibohm:2015twa, Christiansen:2017cxa}. A detailed comparison between the two works therefore requires taking these differences into account. Such a comparison lies beyond the scope of the present work and is left for future work.

Apart from this scheme difference, the two analyses probe slightly different truncations and, taken together, a comprehensive picture emerges: In the present work, we do not include $A^2R$ and $A^4R$ terms. These were considered in \cite{Heisenberg:2026rdk} and found to be irrelevant within that setup. While this evidence is not conclusive, it is at least compatible with our choice to omit these terms in a first exploratory study.
In contrast, we include $\sqrt{g}\,a^3$ contributions, which turn out to be relevant in our FP analysis, whereas they were neglected in \cite{Heisenberg:2026rdk}. This suggests that they should be included in future studies. Finally, the relevance assignment of the $\sqrt{g}\,a^4$ operator differs between the two works: in our case it is irrelevant, while in \cite{Heisenberg:2026rdk} it is relevant. However, we stress that the eigenvectors associated with the critical exponents in \Cref{tab:fixed_points_summary} show sizeable mixing among these operators. Therefore, assigning a given eigenvalue to a single operator should be done with caution. In particular, it remains to be clarified to what extent the omission of $\sqrt{g}\,a^3$ in \cite{Heisenberg:2026rdk} may affect the interpretation of the corresponding eigendirections.

\subsection{Minimally coupled Proca--type fixed point}
\label{sec:MinimallyCoupled} 

The minimally coupled FP exhibits vanishing matter couplings and is therefore the $U(1)$ (Abelian) analogue of the minimally coupled $SU(N)$ (non-Abelian) gauge-theory FP studied in \cite{Christiansen:2017cxa}. It features five relevant directions: two are associated with the gravity sector and form the usual complex-conjugate pair, whose eigendirections are predominantly aligned with the gravitational couplings $g_{h^3}$ and $\mu_h$, with a subleading admixture of $\lambda_{h^3}$. The remaining three lie in the vector sector and are, to a very good approximation, aligned with $\mu_a$, $g_{a^3}$, and $g_{a^4}$, respectively. By contrast, the only irrelevant real direction is predominantly gravitational and mainly aligned with $\lambda_{h^3}$.

This entails that the mass of the vector field can be tuned, and that a specific trajectory approaches the minimally coupled $U(1)$ gauge theory in the physical limit $k\to0$. There, $\mu_a = g_{a^3} = g_{a^4} = 0$, as enforced by the Ward identities. 
All other trajectories correspond to minimally coupled theories with massive vector bosons. In particular, for $\xi\to \infty$, a specific choice may be related to a minimally coupled Proca theory. In the spirit of the programme of scanning asymptotically safe cosmological EFTs, we consider the full set of such theories.

We highlight that the critical exponents of this fixed point are $\xi$-independent, as expected for the minimally coupled Abelian gauge theory, which is obtained along a specific trajectory. This also implies that the number of relevant couplings is the same for all $\xi$. In turn, this may indicate that the remaining theories in this set should rather be seen as massive modifications of the minimally coupled gauge theory.

\subsection{Interacting Proca--type fixed point}
\label{sec:Interacting} 

A further FP is given by the interacting Proca--type branch, which is related to the minimally coupled FP. In contrast to the minimally coupled FP, it features non-vanishing matter self-couplings $\mu_a$ and $g_{a^4}$, which are accompanied by non-minimal matter--gravity couplings $\mu_{a h^n}, g_{a^4 h^n} \propto \mu_a, g_{a^4}$. It exhibits four relevant directions, in contrast to the five of the minimally coupled FP. Two again form the complex-conjugate gravity pair, whose eigendirections are predominantly aligned with the gravitational couplings $g_{h^3}$, $\mu_h$. The remaining two lie in the Proca sector: one is predominantly aligned with $\mu_a$, while the other is mainly aligned with $g_{a^3}$.

Since the couplings $\mu_a$ and $g_{a^3}$ are relevant, one may expect that a specific trajectory leads to an Abelian gauge theory in the physical limit $k\to 0$, with $\mu_a = g_{a^3} = g_{a^4} = 0$. All other trajectories are not related to an Abelian gauge theory. In particular, for $\xi\to \infty$, the interacting FP approaches the minimally coupled one. However, the critical exponents, as well as the number of relevant directions, are different. Hence, we observe two physically distinct FPs lying on top of each other. Moreover, as shown in \Cref{fig:FPs_CritExp}, the critical exponents are approximately constant for $\xi\lesssim 10^{-1}$ and $\xi\gtrsim 10$, with a crossover in between. While this change is sizeable, the number of relevant couplings remains $\xi$-independent.

\subsection{Gemini Proca--type fixed points}
\label{sec:Gemini} 

We find two pairs of conjugate FPs, which we collectively refer to as Gemini$_1$ and Gemini$_2$. Each Gemini pair appears as two branches related by a sign flip of the non-vanishing cubic Proca coupling $g_{a^3}$, while all other couplings take identical values. As a consequence, both FPs within the same pair share the same set of critical exponents. Remarkably, the Gemini FPs are the only ones with a finite fixed-point value of $g_{a^3}$.

Gemini$_1$ features three relevant directions. One of them is a real direction predominantly aligned with the Proca mass parameter $\mu_a$. The remaining two form the standard complex-conjugate pair in the gravitational sector. While the counting of relevant directions is robust, the corresponding eigendirections exhibit substantial overlap between the gravity and vector sectors, so that a sharp operator-by-operator assignment is less clear--cut than for the minimally coupled or interacting branches. This pair exists for $\xi > \xi^{\text{Gemini}_1}_{\rm min} \approx 5$, and its critical exponents depend only mildly on $\xi$, while the fixed-point values show a stronger $\xi$ dependence.

In contrast, Gemini$_2$ features five relevant directions. Two of them are associated with the gravity sector and form the usual complex-conjugate pair. Among the remaining three, the real direction is mainly aligned with $\mu_a$, while the other two form a complex-conjugate pair. The only irrelevant real direction is mostly gravitational and clearly aligned with $\lambda_{h^3}$. This pair exists for $\xi > \xi^{\text{Gemini}_2}_{\rm min} \approx 25$. Both the critical exponents and the fixed-point values depend only mildly on $\xi$.

For $\xi\to\infty$, the fixed-point values in the Proca sector of both Gemini$_1$ and Gemini$_2$ approach those of the minimally coupled Proca--type FP. We note, however, that the FPs do not merge in this limit, as they remain distinguished by both the number of relevant directions and their critical exponents. The $\xi$-dependence of all fixed-point values and critical exponents is shown in \Cref{fig:FPs_CritExp,fig:ProcaFPscan_all_withGemini}.

\subsection{Interacting Proca$^\star$ fixed point}
\label{sec:Isolated} 

Within the full truncation and in the strict Proca limit $\xi\to\infty$, we find an interacting FP with four relevant directions. Two are associated with the gravity sector and form the usual complex-conjugate pair predominantly aligned with $g_{h^3}$ and $\mu_h$. The remaining two lie in the Proca sector: one is exactly aligned with $g_{a^3}$, while the other is predominantly aligned with $\mu_a$.

The disappearance of this FP for $\xi\lesssim 4$ indicates that it is a genuine vector-theory FP and that no UV--IR trajectory leads to an Abelian gauge theory in the physical limit $k\to 0$. Hence, it is a genuine interacting Proca--type FP, which we denote by \textit{interacting Proca$^\star$}. We also emphasise that it is the only FP whose matter self-couplings do not vanish identically in the limit $\xi\to \infty$. Moreover, the relatively small critical exponents and the small number of relevant directions suggest that this FP may remain stable under systematic improvements of the approximation.

\section{Conclusions}
\label{sec:Conclusions}

In this work, we have taken a first step towards the broad programme of connecting cosmology--driven EFTs with the question of whether they admit a consistent UV completion within QFT. Motivated by the fact that an EFT may successfully describe late--time dynamics and still break down at high energies, we view UV consistency as a selection principle that can narrow the space of viable modified-gravity scenarios and sharpen their IR predictions. With this aim, we initiate a systematic exploration of the UV--consistent landscape of cosmological EFTs within the AS paradigm, using the non--perturbative fRG and focusing here on V--T theories of the generalised Proca type.

We have analysed FP solutions of a V--T subsector of GPTs, employing a nontrivial truncation that includes higher-dimensional operators generated by a vector condensate. To interpolate between gauge-like and Proca dynamics, we have introduced a gauge-fixing--like regularisation of the longitudinal vector mode.

We have identified several FPs in the coupled gravity--Proca system that are smoothly connected to the pure-gravity Reuter FP. In particular, we find a remarkable interacting Proca--type FP with four relevant directions. These align with two directions in the gravity sector, matching those of the Reuter FP, and two arising from the Proca sector. This solution provides evidence for non--perturbative renormalisability within our truncation and defines a UV starting point for RG trajectories flowing towards phenomenologically viable IR Proca theories, thereby constraining their parameter space. We also find two pairs of conjugate solutions, denoted Gemini$_1$ and Gemini$_2$, characterised by a non-vanishing cubic Proca coupling and a complex-conjugate pair of critical exponents in the Proca sector.

From the cosmological point of view, the central outcome of the present study is that UV consistency in the gravity--Proca system can act as a concrete selection criterion on the parameter space of late--time EFTs. The finite number of relevant directions at the interacting Proca--type FP implies that only a restricted set of RG trajectories emanate from the UV, and these trajectories encode correlations among the IR couplings that enter cosmological observables. This provides a pathway to translate the UV critical surface into constraints on the EFT data used in cosmological analyses, thereby reducing the available parameter space compared to a purely bottom-up EFT approach.

Beyond the scope of this study, there are several directions in which our work can and will be extended. On the technical side, it will be important to enlarge the truncation by including the marginal operators that we have set to zero here, in particular the $A^2R$ interaction and the curvature-square operators, and to incorporate additional momentum--dependent information beyond our approximation. On the physics side, the UV FPs identified here provide UV initial conditions for RG trajectories that can be evolved towards the IR and translated into the EFT language used in cosmology. This will allow one to combine UV consistency with standard viability criteria for cosmological perturbations and with observational bounds, and thus to make the UV-consistency selection principle proposed here quantitative.

Taken together, our results provide evidence that cosmologically motivated Proca--type EFTs can admit an AS UV completion when coupled to gravity, supporting the broader perspective that UV consistency can sharpen the predictive power of cosmological EFTs and guide the search for viable extensions of the cosmological SM. We expect that extending the present analysis along the lines outlined above will further clarify the interplay between UV FP structure, EFT parameter correlations, and observational viability, and will help clarify to what extent AS V--T theories can provide a minimal and predictive framework for late--time cosmology.


\begin{acknowledgments}
This work is funded by the Deutsche Forschungsgemeinschaft (DFG, German Research Foundation) under Germany’s Excellence Strategy EXC 2181/1 - 390900948 (the Heidelberg STRUCTURES Excellence Cluster) and the Collaborative Research Centre SFB 1225 - 273811115 (ISOQUANT). APG is supported by the RIKEN Special Postdoctoral Researcher (SPDR) Program. MR acknowledges support by the Science and Technology Facilities Council under the Consolidated Grant ST/X000796/1 and the Ernest Rutherford Fellowship ST/Z510282/1.
\end{acknowledgments}

\section*{Appendices}
\crefalias{section}{appsec}
\appendix
\begin{figure*}[t!]
  \centering
  \includegraphics[width=0.7\textwidth, 
  clip, trim=0pt 0pt 25pt 0pt 
  ]{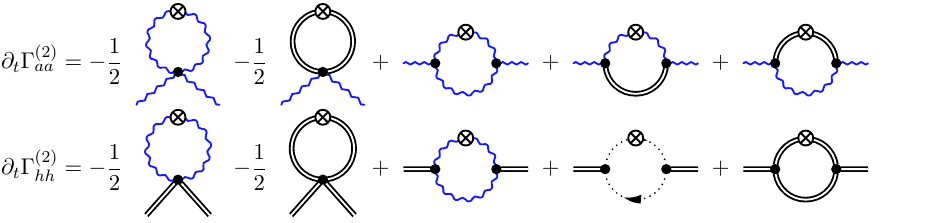}
  \caption{Diagrammatic flow equations for the two-point functions in the gravity--Proca system. Wiggly blue lines represent Proca fluctuating fields, double lines gravitons, and directed dashed lines ghosts. Filled dots indicate interaction vertices and `$\otimes$' denotes the regulator insertion $\partial_t R_k$.}
  \label{fig:Flow2p}
\end{figure*}
%

\section{Proca vertices and regulator insertion}
\label{app:VerticesAndTensorStructures}

In this appendix, we summarise the Proca building blocks entering our fRG computations, namely the $n$-point functions, propagators, and regulator insertions. All expressions have been computed by perturbing the action \labelcref{eq:GammaApprox} with \labelcref{eq:GFunExpansion} according to \labelcref{eq:MetricExpansion} using \texttt{VertEXpand} \cite{vertexpand}, and are presented in Euclidean signature at the momentum-symmetric point. The diagrammatic expressions were derived with DoFun \cite{Huber:2019dkb}, and the contractions were carried out with FormTracer \cite{Cyrol:2016zqb} and with the help of self-written codes that are based on FORM \cite{Vermaseren:2000nd, Kuipers:2012rf, Ruijl:2017dtg, Davies:2026cci}. An overview of codes used within the AS community is presented in the AS codebase \cite{AScodebase}.

In the graviton sector, we do not list explicit vertices, as within our truncation they are identical to the corresponding pure-gravity expressions, with the same canonical tensor structures as in the Einstein--Hilbert case and can be found in \cite{Denz:2016qks}. The Proca $n$-point functions $\Gamma^{(n)}_{a\cdots a}$, expressed in terms of the dimensionless couplings in \labelcref{eq:boldg-FluctuationApproach}, are given by
\begin{eqnarray}\label{eq:npointProca_case1_V2}
    \Big[\Gamma^{(2)}_{aa}\Big]_{\mu\nu} & = & Z_a\left\{k^2\mu_a\mathcal{T}_{1,\,\mu\nu}^{(aa)}+ \mathcal{T}_{2,\,\mu\nu}^{(aa)}\right\}, \nonumber\\[1ex]
    \Big[\Gamma^{(3)}_{aaa}\Big]_{\mu\nu\rho} & = & i\,Z_a^{3/2}g_{a^3} \mathcal{T}_{1,\,\mu\nu\rho}^{(3a)},  \nonumber\\[1ex]
    \Big[\Gamma^{(4)}_{aaaa}\Big]_{\mu\nu\rho\sigma} & = & \,Z_a^2g_{a^4}\mathcal{T}_{1,\,\mu\nu\rho\sigma}^{(4a)},
\end{eqnarray}
with tensor structures
\begin{align}
    \mathcal{T}_{1,\,\mu\nu}^{(aa)}&=\delta_{\mu\nu}, \nonumber\\[1ex]
    \mathcal{T}_{2,\,\mu\nu}^{(aa)}&=\delta_{\mu\nu}p^2-p_{\mu}p_{\nu}, \nonumber \\[1ex]
    \mathcal{T}_{1,\,\mu\nu\rho}^{(3a)}&=2\left[\delta_{\nu\rho}\,p_{1,\mu}+\delta_{\mu\rho}\,p_{2,\nu}-\delta_{\mu\nu}(p_{1,\rho}+p_{2,\rho})\right], \nonumber\\[1ex]
    \mathcal{T}_{1,\,\mu\nu\rho\sigma}^{(4a)}&=2\left[\delta_{\nu\rho}\delta_{\mu\sigma}+\delta_{\mu\rho}\delta_{\nu\sigma}+\delta_{\mu\nu}\delta_{\rho\sigma}\right].
    \label{eq:ProcaTensorStructures}
\end{align}
From inverting the two-point function $\Gamma^{(2)}_{aa}$, together with the pseudo--gauge fixing term, we obtain the corresponding propagator in \labelcref{eq:Procaaa}. Alternatively, one can decompose it into transverse/longitudinal modes using the projectors,
\begin{align}
  \Pi^{\mu\nu}_{\mathrm{T}}(p) &= \delta^{\mu\nu}-\frac{p^\mu p^\nu}{p^2}, &
  \Pi^{\mu\nu}_{\mathrm{L}}(p) &= \frac{p^\mu p^\nu}{p^2},
\end{align}
and use the Litim-type shape function \cite{Litim:2001up}
\begin{align}
  r_k(x)=\Big(\frac{1}{x}-1\Big)\,\Theta(1-x), \qquad x=\frac{p^2}{k^2}.
\end{align}
The regulated propagator reads
\begin{align}\label{eq:ProcaPropagator_decomp}
    G^{\mu\nu}_{aa}(p) = \frac{1}{Z_a}\Bigg(&\frac{\Pi^{\mu\nu}_{\mathrm{T}}(p)}{\,p^{2}\left[1+r_k\left(x\right)\right] + k^2\mu_a}\nonumber\\
    &+\frac{\Pi^{\mu\nu}_{\mathrm{L}}(p)}{\,(1/\xi)\,p^{2}\left[1+r_k\left(x\right)\right] + k^2\mu_a}\Bigg),
\end{align}
where $\xi$ is the Proca pseudo--gauge fixing parameter \cite{Jung:2019nnr}. The regulator insertion is chosen as
\begin{align}
    R_k^{\mu\nu}(p)= Z_a\left[\Pi^{\mu\nu}_{\mathrm{T}}(p)+ \frac1{\xi}\,\Pi^{\mu\nu}_{\mathrm{L}}(p)\right]\,p^{2}\,r_k\left(x\right),
\label{eq:ProcaReg}
\end{align}
so that its RG--time derivative entering the flow is given by
\begin{align}
    \partial_t R_k^{\mu\nu} =& Z_a\left(\Pi^{\mu\nu}_{\rm T}+\frac1{\xi}\,\Pi^{\mu\nu}_{\rm L} \right)\,p^{2}\,\partial_t r_k\left(x\right)-\eta_a\,R_k^{\mu\nu}\nonumber\\[1ex]
    =& Z_a\left(\Pi^{\mu\nu}_{\rm T}+\frac1{\xi}\Pi^{\mu\nu}_{\rm L}\right)\,p^{2}\left[\frac{2k^{2}}{p^{2}}-\eta_a\left(\frac{k^{2}}{p^{2}}-1\right)\right]\nonumber\\[1ex]
    &\times\Theta\!\left(1-\frac{p^{2}}{k^{2}}\right),
\label{eq:dRProca_Litim}
\end{align}
where we have already used that the term proportional to the delta function does not contribute in our computation.

\begin{figure*}[tbp]
  \centering
  \makebox[\textwidth]{
  \begin{minipage}{0.49\textwidth}
    \centering
        \includegraphics[width=0.90\columnwidth, clip, trim=59pt 0pt 63pt 0pt]{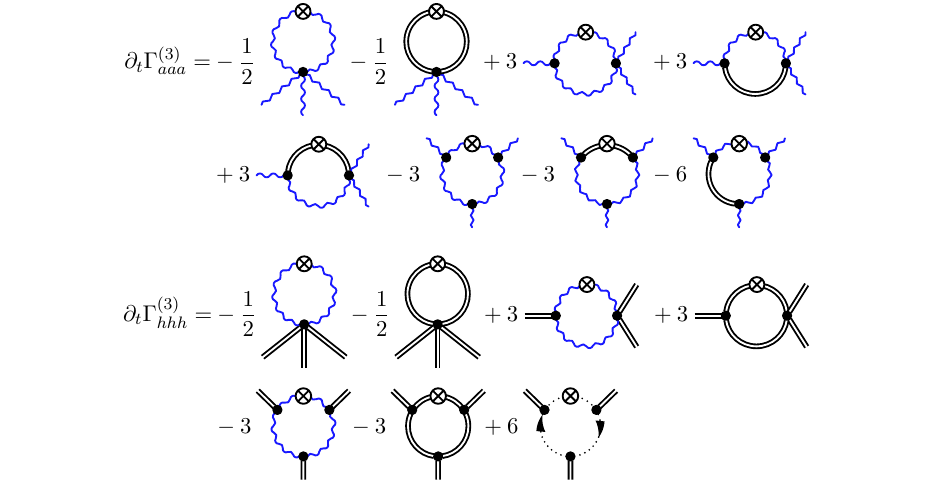}
    \end{minipage}\hfill
    \begin{minipage}{0.49\textwidth}
    \centering
        \includegraphics[width=0.99\columnwidth, clip, trim=35pt 0pt 36pt 0pt]{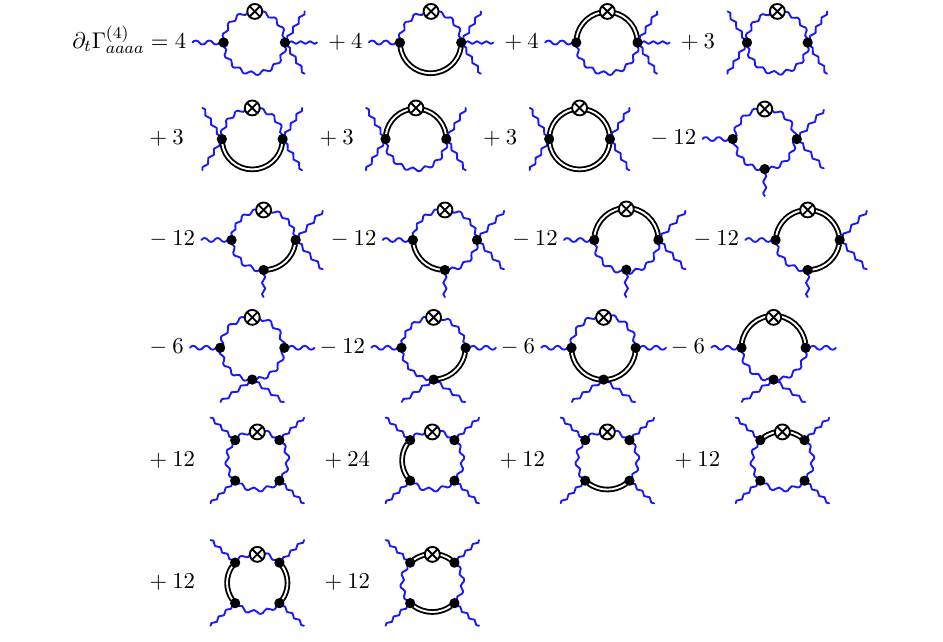}
    \end{minipage}
    }
    \caption{Diagrammatic flow equations for the three- (left) and four- (right) point functions in the gravity--Proca system; line and symbol conventions are the same as in \Cref{fig:Flow2p}.}
  \label{fig:Flow3p4p}
\end{figure*}
%

\section{Projected flow equations}
\label{app:ProjectedFlowEquations}
\crefalias{section}{appsec}
We compute the flow of $n$-point correlation functions, which are derived from \labelcref{eq:FlowEffAct} by taking appropriate functional derivatives with respect to the fluctuation fields, subsequently evaluating the flow at symmetric external kinematics and projecting onto tensor structures of our choice. The resulting RG flows for the V--T system are expanded in a derivative expansion around vanishing external momenta. Note that the derivative expansion works well for the Newton coupling but is not able to capture the momentum dependence of the graviton anomalous dimension, see \cite{Christiansen:2014raa, Christiansen:2015rva, Pawlowski:2020qer, Pawlowski:2023gym}. Therefore, we use $\eta_h=\eta_c=0$ for the sake of simplicity and only consider a non-vanishing $\eta_a$ in the canonical terms of the beta functions, denoted as $\eta_{a,0}$, while we set $\eta_a=0$ in the loops.

For the flows of the graviton and ghost sectors, we proceed as in \cite{Christiansen:2015rva,Denz:2016qks}: We work in the harmonic Landau gauge, $\beta = 1$ and $\alpha\to 0$, and project the graviton two-point function onto its transverse-traceless (TT) component to extract $\mu_h$. We extract $g_{h^3}$ and $\lambda_{h^3}$ from TT-projected three-graviton vertices using, respectively, a projection onto the Ricci tensor structure and onto the cosmological-constant tensor structure. Ghost flows are obtained from a single $\delta_{\mu\nu}$ projection. In the Proca sector, projections onto the corresponding tensor structures isolate the mass and anomalous dimension from $\Gamma^{(2)}_{aa}$, and the cubic/quartic self--interactions from $\Gamma^{(3)}_{aaa}$/$\Gamma^{(4)}_{aaaa}$. These projections are listed below.

\subsection*{Flow of $\Gamma^{(2)}_{aa}$:}
We isolate $\eta_a$ and the flow of $\mu_a$ when projecting $\Gamma^{(2)}_{aa}$ in the following way:
\begin{align}
    \mathcal{P}_{1}^{(aa)\,\mu\nu}\times \left[\Gamma^{(2)}_{aa}\right]_{\mu\nu}\Big|_{p=0} &= Z_a\,k^2\mu_a,\\
    \mathcal{P}_{2}^{(aa)\,\mu\nu}\times \partial_{p^2}\left[\Gamma^{(2)}_{aa}\right]_{\mu\nu}\Big|_{p=0}&=Z_a;
\end{align}
where we introduced the projectors
\begin{align}
    \mathcal{P}_{1,\,\mu\nu}^{(aa)}=\frac{\mathcal{T}^{(aa)}_{1,\,\mu\nu}}{\mathrm{Tr}^{(aa)}_{11}}, \qquad
    \mathcal{P}_{2,\,\mu\nu}^{(aa)}=\frac{\mathcal{T}^{(aa)}_{2,\,\mu\nu}}{\mathrm{Tr}^{(aa)}_{12}};
\end{align}
and make use of the compact notation $\mathrm{Tr}^{(aa)}_{ij}\equiv \mathrm{Tr}\{\mathcal{T}^{(aa)\,\mu\nu}_i\mathcal{T}^{(aa)}_{j,\,\mu\nu}\}$. When computing the scale derivative, the terms $\partial_t \Gamma^{(2)}_{aa}$ are what we refer to as the flow of the Proca two-point function, or $\mathrm{Flow}[\Gamma^{(2)}_{aa}]$. Solving the system, the Proca anomalous dimension and beta function \labelcref{eq:FP} for $\mu_a$ read:
\begin{align}
    \partial_t\mu_a &= \mu_a(\eta_{a,0}-2) + \frac{\mathcal{P}_1^{(aa)}\mathrm{Flow}\big[\Gamma^{(2)}_{aa}\big]}{k^2Z_a}\bigg|_{p=0}, \label{eq:Flow_muA}\\[1ex]
    \eta_a &= -\frac{\mathcal{P}_2^{(aa)}\partial_{p^2}\mathrm{Flow}\big[\Gamma^{(2)}_{aa}\big]}{Z_a}\bigg|_{p=0}. 
    \label{eq:Flow_etaA}
\end{align}
%

\subsection*{Flow of $\Gamma^{(3)}_{aaa}$:}
We project the three-point function at $p_3=-p_1-p_2$, as imposed by momentum conservation, with
\begin{align}
    \mathcal{P}^{(3a)}_{1,\,\mu\nu\rho} = \frac{\mathcal{T}^{(3a)}_{1,\,\mu\nu\rho}}{\mathrm{Tr}^{(3a)}_{11}},
\end{align}
so that, after taking the scale derivative,
\begin{align}
    \partial_t g_{a^3} = \frac{3}{2}\eta_{a,0}\,g_{a^3}- \frac{i\,\mathcal{P}_1^{(3a)}\,\mathrm{Flow}\big[\Gamma^{(3)}_{aaa}\big]}{Z_a^{3/2}}\Bigg|_{p=0}.
\end{align}
%

\subsection*{Flow of $\Gamma^{(4)}_{aaaa}$:}
For the 4-point function, we take $p_4=-(p_1+p_2+p_3)$ and use the projector:
\begin{align}
    \mathcal{P}^{(4a)}_{1,\,\mu\nu\rho\sigma} = \frac{\mathcal{T}^{(4a)}_{1,\,\mu\nu\rho\sigma}}{\mathrm{Tr}^{(4a)}_{11}},
\end{align}
which isolates $g_{a^4}$. Its flow then reads:
\begin{align}
    \partial_tg_{a^4}&= 2\eta_{a,0}\,g_{a^4}+ \frac{\mathcal{P}_1^{(4a)}\,\mathrm{Flow}\big[\Gamma^{(4)}_{aaaa}\big]}{Z_a^2}\Bigg|_{p=0}.
\end{align}

\clearpage
\bibliography{Proca_AS.bib}

\begin{thebibliography}{44}%
\makeatletter
\providecommand \@ifxundefined [1]{%
 \@ifx{#1\undefined}
}%
\providecommand \@ifnum [1]{%
 \ifnum #1\expandafter \@firstoftwo
 \else \expandafter \@secondoftwo
 \fi
}%
\providecommand \@ifx [1]{%
 \ifx #1\expandafter \@firstoftwo
 \else \expandafter \@secondoftwo
 \fi
}%
\providecommand \natexlab [1]{#1}%
\providecommand \enquote  [1]{``#1''}%
\providecommand \bibnamefont  [1]{#1}%
\providecommand \bibfnamefont [1]{#1}%
\providecommand \citenamefont [1]{#1}%
\providecommand \href@noop [0]{\@secondoftwo}%
\providecommand \href [0]{\begingroup \@sanitize@url \@href}%
\providecommand \@href[1]{\@@startlink{#1}\@@href}%
\providecommand \@@href[1]{\endgroup#1\@@endlink}%
\providecommand \@sanitize@url [0]{\catcode `\\12\catcode `\$12\catcode `\&12\catcode `\#12\catcode `\^12\catcode `\_12\catcode `\%12\relax}%
\providecommand \@@startlink[1]{}%
\providecommand \@@endlink[0]{}%
\providecommand \url  [0]{\begingroup\@sanitize@url \@url }%
\providecommand \@url [1]{\endgroup\@href {#1}{\urlprefix }}%
\providecommand \urlprefix  [0]{URL }%
\providecommand \Eprint [0]{\href }%
\providecommand \doibase [0]{https://doi.org/}%
\providecommand \selectlanguage [0]{\@gobble}%
\providecommand \bibinfo  [0]{\@secondoftwo}%
\providecommand \bibfield  [0]{\@secondoftwo}%
\providecommand \translation [1]{[#1]}%
\providecommand \BibitemOpen [0]{}%
\providecommand \bibitemStop [0]{}%
\providecommand \bibitemNoStop [0]{.\EOS\space}%
\providecommand \EOS [0]{\spacefactor3000\relax}%
\providecommand \BibitemShut  [1]{\csname bibitem#1\endcsname}%
\let\auto@bib@innerbib\@empty
\bibitem [{\citenamefont {Heisenberg}(2019)}]{Heisenberg:2018vsk}%
  \BibitemOpen
  \bibfield  {author} {\bibinfo {author} {\bibfnamefont {L.}~\bibnamefont {Heisenberg}},\ }\bibfield  {title} {\bibinfo {title} {{A systematic approach to generalisations of General Relativity and their cosmological implications}},\ }\href {https://doi.org/10.1016/j.physrep.2018.11.006} {\bibfield  {journal} {\bibinfo  {journal} {Phys. Rept.}\ }\textbf {\bibinfo {volume} {796}},\ \bibinfo {pages} {1} (\bibinfo {year} {2019})},\ \Eprint {https://arxiv.org/abs/1807.01725} {arXiv:1807.01725 [gr-qc]} \BibitemShut {NoStop}%
\bibitem [{\citenamefont {Bambi}\ \emph {et~al.}(2024)\citenamefont {Bambi}, \citenamefont {Modesto},\ and\ \citenamefont {Shapiro}}]{Bambi:2023jiz}%
  \BibitemOpen
  \bibinfo {editor} {\bibfnamefont {C.}~\bibnamefont {Bambi}}, \bibinfo {editor} {\bibfnamefont {L.}~\bibnamefont {Modesto}},\ and\ \bibinfo {editor} {\bibfnamefont {I.}~\bibnamefont {Shapiro}},\ eds.,\ \href {https://doi.org/10.1007/978-981-99-7681-2} {\emph {\bibinfo {title} {{Handbook of Quantum Gravity}}}}\ (\bibinfo  {publisher} {Springer},\ \bibinfo {year} {2024})\BibitemShut {NoStop}%
\bibitem [{\citenamefont {Bonanno}\ \emph {et~al.}(2020)\citenamefont {Bonanno}, \citenamefont {Eichhorn}, \citenamefont {Gies}, \citenamefont {Pawlowski}, \citenamefont {Percacci}, \citenamefont {Reuter}, \citenamefont {Saueressig},\ and\ \citenamefont {Vacca}}]{Bonanno:2020bil}%
  \BibitemOpen
  \bibfield  {author} {\bibinfo {author} {\bibfnamefont {A.}~\bibnamefont {Bonanno}}, \bibinfo {author} {\bibfnamefont {A.}~\bibnamefont {Eichhorn}}, \bibinfo {author} {\bibfnamefont {H.}~\bibnamefont {Gies}}, \bibinfo {author} {\bibfnamefont {J.~M.}\ \bibnamefont {Pawlowski}}, \bibinfo {author} {\bibfnamefont {R.}~\bibnamefont {Percacci}}, \bibinfo {author} {\bibfnamefont {M.}~\bibnamefont {Reuter}}, \bibinfo {author} {\bibfnamefont {F.}~\bibnamefont {Saueressig}},\ and\ \bibinfo {author} {\bibfnamefont {G.~P.}\ \bibnamefont {Vacca}},\ }\bibfield  {title} {\bibinfo {title} {{Critical reflections on asymptotically safe gravity}},\ }\href {https://doi.org/10.3389/fphy.2020.00269} {\bibfield  {journal} {\bibinfo  {journal} {Front. in Phys.}\ }\textbf {\bibinfo {volume} {8}},\ \bibinfo {pages} {269} (\bibinfo {year} {2020})},\ \Eprint {https://arxiv.org/abs/2004.06810} {arXiv:2004.06810 [gr-qc]} \BibitemShut {NoStop}%
\bibitem [{\citenamefont {Dupuis}\ \emph {et~al.}(2021)\citenamefont {Dupuis}, \citenamefont {Canet}, \citenamefont {Eichhorn}, \citenamefont {Metzner}, \citenamefont {Pawlowski}, \citenamefont {Tissier},\ and\ \citenamefont {Wschebor}}]{Dupuis:2020fhh}%
  \BibitemOpen
  \bibfield  {author} {\bibinfo {author} {\bibfnamefont {N.}~\bibnamefont {Dupuis}}, \bibinfo {author} {\bibfnamefont {L.}~\bibnamefont {Canet}}, \bibinfo {author} {\bibfnamefont {A.}~\bibnamefont {Eichhorn}}, \bibinfo {author} {\bibfnamefont {W.}~\bibnamefont {Metzner}}, \bibinfo {author} {\bibfnamefont {J.~M.}\ \bibnamefont {Pawlowski}}, \bibinfo {author} {\bibfnamefont {M.}~\bibnamefont {Tissier}},\ and\ \bibinfo {author} {\bibfnamefont {N.}~\bibnamefont {Wschebor}},\ }\bibfield  {title} {\bibinfo {title} {{The nonperturbative functional renormalization group and its applications}},\ }\href {https://doi.org/10.1016/j.physrep.2021.01.001} {\bibfield  {journal} {\bibinfo  {journal} {Phys. Rept.}\ }\textbf {\bibinfo {volume} {910}},\ \bibinfo {pages} {1} (\bibinfo {year} {2021})},\ \Eprint {https://arxiv.org/abs/2006.04853} {arXiv:2006.04853 [cond-mat.stat-mech]} \BibitemShut {NoStop}%
\bibitem [{\citenamefont {Knorr}\ \emph {et~al.}(2023)\citenamefont {Knorr}, \citenamefont {Ripken},\ and\ \citenamefont {Saueressig}}]{Knorr:2022dsx}%
  \BibitemOpen
  \bibfield  {author} {\bibinfo {author} {\bibfnamefont {B.}~\bibnamefont {Knorr}}, \bibinfo {author} {\bibfnamefont {C.}~\bibnamefont {Ripken}},\ and\ \bibinfo {author} {\bibfnamefont {F.}~\bibnamefont {Saueressig}},\ }\bibfield  {title} {\bibinfo {title} {{Form Factors in Asymptotically Safe Quantum Gravity}},\ }in\ \href {https://doi.org/10.1007/978-981-19-3079-9_21-1} {\emph {\bibinfo {booktitle} {Handbook of Quantum Gravity}}}\ (\bibinfo  {publisher} {Springer Nature Singapore},\ \bibinfo {address} {Singapore},\ \bibinfo {year} {2023})\ \Eprint {https://arxiv.org/abs/2210.16072} {arXiv:2210.16072 [hep-th]} \BibitemShut {NoStop}%
\bibitem [{\citenamefont {Eichhorn}\ and\ \citenamefont {Schiffer}(2022)}]{Eichhorn:2022gku}%
  \BibitemOpen
  \bibfield  {author} {\bibinfo {author} {\bibfnamefont {A.}~\bibnamefont {Eichhorn}}\ and\ \bibinfo {author} {\bibfnamefont {M.}~\bibnamefont {Schiffer}},\ }\bibfield  {title} {\bibinfo {title} {{Asymptotic safety of gravity with matter}},\ }\href@noop {} {\  (\bibinfo {year} {2022})},\ \Eprint {https://arxiv.org/abs/2212.07456} {arXiv:2212.07456 [hep-th]} \BibitemShut {NoStop}%
\bibitem [{\citenamefont {Morris}\ and\ \citenamefont {Stulga}(2023)}]{Morris:2022btf}%
  \BibitemOpen
  \bibfield  {author} {\bibinfo {author} {\bibfnamefont {T.~R.}\ \bibnamefont {Morris}}\ and\ \bibinfo {author} {\bibfnamefont {D.}~\bibnamefont {Stulga}},\ }\bibinfo {title} {{The Functional f(R) Approximation}}\ (\bibinfo {year} {2023})\ \Eprint {https://arxiv.org/abs/2210.11356} {arXiv:2210.11356 [hep-th]} \BibitemShut {NoStop}%
\bibitem [{\citenamefont {Wetterich}(2023)}]{Wetterich:2022ncl}%
  \BibitemOpen
  \bibfield  {author} {\bibinfo {author} {\bibfnamefont {C.}~\bibnamefont {Wetterich}},\ }\bibfield  {title} {\bibinfo {title} {{Quantum Gravity and Scale Symmetry in Cosmology}},\ }in\ \href {https://doi.org/10.1007/978-981-19-3079-9_26-1} {\emph {\bibinfo {booktitle} {Handbook of Quantum Gravity}}}\ (\bibinfo  {publisher} {Springer Nature Singapore},\ \bibinfo {address} {Singapore},\ \bibinfo {year} {2023})\ \Eprint {https://arxiv.org/abs/2211.03596} {arXiv:2211.03596 [gr-qc]} \BibitemShut {NoStop}%
\bibitem [{\citenamefont {Martini}\ \emph {et~al.}(2023)\citenamefont {Martini}, \citenamefont {Vacca},\ and\ \citenamefont {Zanusso}}]{Martini:2022sll}%
  \BibitemOpen
  \bibfield  {author} {\bibinfo {author} {\bibfnamefont {R.}~\bibnamefont {Martini}}, \bibinfo {author} {\bibfnamefont {G.~P.}\ \bibnamefont {Vacca}},\ and\ \bibinfo {author} {\bibfnamefont {O.}~\bibnamefont {Zanusso}},\ }\bibfield  {title} {\bibinfo {title} {{Perturbative Approaches to Nonperturbative Quantum Gravity}},\ }in\ \href {https://doi.org/10.1007/978-981-19-3079-9_25-1} {\emph {\bibinfo {booktitle} {Handbook of Quantum Gravity}}}\ (\bibinfo  {publisher} {Springer Nature Singapore},\ \bibinfo {address} {Singapore},\ \bibinfo {year} {2023})\ pp.\ \bibinfo {pages} {1--46},\ \Eprint {https://arxiv.org/abs/2210.13910} {arXiv:2210.13910 [hep-th]} \BibitemShut {NoStop}%
\bibitem [{\citenamefont {Saueressig}(2023)}]{Saueressig:2023irs}%
  \BibitemOpen
  \bibfield  {author} {\bibinfo {author} {\bibfnamefont {F.}~\bibnamefont {Saueressig}},\ }\bibfield  {title} {\bibinfo {title} {{The Functional Renormalization Group in Quantum Gravity}},\ }in\ \href {https://doi.org/10.1007/978-981-19-3079-9_16-1} {\emph {\bibinfo {booktitle} {Handbook of Quantum Gravity}}}\ (\bibinfo  {publisher} {Springer Nature Singapore},\ \bibinfo {address} {Singapore},\ \bibinfo {year} {2023})\ pp.\ \bibinfo {pages} {1--33},\ \Eprint {https://arxiv.org/abs/2302.14152} {arXiv:2302.14152 [hep-th]} \BibitemShut {NoStop}%
\bibitem [{\citenamefont {Pawlowski}\ and\ \citenamefont {Reichert}(2023)}]{Pawlowski:2023gym}%
  \BibitemOpen
  \bibfield  {author} {\bibinfo {author} {\bibfnamefont {J.~M.}\ \bibnamefont {Pawlowski}}\ and\ \bibinfo {author} {\bibfnamefont {M.}~\bibnamefont {Reichert}},\ }\bibfield  {title} {\bibinfo {title} {{Quantum Gravity from dynamical metric fluctuations}},\ }in\ \href {https://doi.org/10.1007/978-981-19-3079-9_17-1} {\emph {\bibinfo {booktitle} {Handbook of Quantum Gravity}}}\ (\bibinfo  {publisher} {Springer Nature Singapore},\ \bibinfo {address} {Singapore},\ \bibinfo {year} {2023})\ \Eprint {https://arxiv.org/abs/2309.10785} {arXiv:2309.10785 [hep-th]} \BibitemShut {NoStop}%
\bibitem [{\citenamefont {Platania}(2023)}]{Platania:2023srt}%
  \BibitemOpen
  \bibfield  {author} {\bibinfo {author} {\bibfnamefont {A.}~\bibnamefont {Platania}},\ }\bibfield  {title} {\bibinfo {title} {Black holes in asymptotically safe gravity},\ }in\ \href {https://doi.org/10.1007/978-981-19-3079-9_24-1} {\emph {\bibinfo {booktitle} {Handbook of Quantum Gravity}}}\ (\bibinfo  {publisher} {Springer Nature Singapore},\ \bibinfo {address} {Singapore},\ \bibinfo {year} {2023})\ \Eprint {https://arxiv.org/abs/2302.04272} {arXiv:2302.04272 [gr-qc]} \BibitemShut {NoStop}%
\bibitem [{\citenamefont {Bonanno}(2023)}]{Bonanno:2024xne}%
  \BibitemOpen
  \bibfield  {author} {\bibinfo {author} {\bibfnamefont {A.}~\bibnamefont {Bonanno}},\ }\bibfield  {title} {\bibinfo {title} {{Asymptotic Safety and Cosmology}},\ }in\ \href {https://doi.org/10.1007/978-981-19-3079-9_23-1} {\emph {\bibinfo {booktitle} {Handbook of Quantum Gravity}}}\ (\bibinfo  {publisher} {Springer Nature Singapore},\ \bibinfo {address} {Singapore},\ \bibinfo {year} {2023})\BibitemShut {NoStop}%
\bibitem [{\citenamefont {Reichert}(2020)}]{Reichert:2020mja}%
  \BibitemOpen
  \bibfield  {author} {\bibinfo {author} {\bibfnamefont {M.}~\bibnamefont {Reichert}},\ }\bibfield  {title} {\bibinfo {title} {{Lecture notes: Functional Renormalisation Group and Asymptotically Safe Quantum Gravity}},\ }\href {https://doi.org/10.22323/1.384.0005} {\bibfield  {journal} {\bibinfo  {journal} {PoS}\ }\textbf {\bibinfo {volume} {384}},\ \bibinfo {pages} {005} (\bibinfo {year} {2020})}\BibitemShut {NoStop}%
\bibitem [{\citenamefont {Basile}\ \emph {et~al.}(2025)\citenamefont {Basile}, \citenamefont {Buoninfante}, \citenamefont {Di~Filippo}, \citenamefont {Knorr}, \citenamefont {Platania},\ and\ \citenamefont {Tokareva}}]{Basile:2024oms}%
  \BibitemOpen
  \bibfield  {author} {\bibinfo {author} {\bibfnamefont {I.}~\bibnamefont {Basile}}, \bibinfo {author} {\bibfnamefont {L.}~\bibnamefont {Buoninfante}}, \bibinfo {author} {\bibfnamefont {F.}~\bibnamefont {Di~Filippo}}, \bibinfo {author} {\bibfnamefont {B.}~\bibnamefont {Knorr}}, \bibinfo {author} {\bibfnamefont {A.}~\bibnamefont {Platania}},\ and\ \bibinfo {author} {\bibfnamefont {A.}~\bibnamefont {Tokareva}},\ }\bibfield  {title} {\bibinfo {title} {{Lectures in quantum gravity}},\ }\href {https://doi.org/10.21468/SciPostPhysLectNotes.98} {\bibfield  {journal} {\bibinfo  {journal} {SciPost Phys. Lect. Notes}\ }\textbf {\bibinfo {volume} {98}},\ \bibinfo {pages} {1} (\bibinfo {year} {2025})},\ \Eprint {https://arxiv.org/abs/2412.08690} {arXiv:2412.08690 [hep-th]} \BibitemShut {NoStop}%
\bibitem [{\citenamefont {Eichhorn}\ \emph {et~al.}(2025)\citenamefont {Eichhorn}, \citenamefont {Hebecker}, \citenamefont {Pawlowski},\ and\ \citenamefont {Walcher}}]{Eichhorn:2024rkc}%
  \BibitemOpen
  \bibfield  {author} {\bibinfo {author} {\bibfnamefont {A.}~\bibnamefont {Eichhorn}}, \bibinfo {author} {\bibfnamefont {A.}~\bibnamefont {Hebecker}}, \bibinfo {author} {\bibfnamefont {J.~M.}\ \bibnamefont {Pawlowski}},\ and\ \bibinfo {author} {\bibfnamefont {J.}~\bibnamefont {Walcher}},\ }\bibfield  {title} {\bibinfo {title} {{The absolute swampland}},\ }\href {https://doi.org/10.1209/0295-5075/ada1f3} {\bibfield  {journal} {\bibinfo  {journal} {EPL}\ }\textbf {\bibinfo {volume} {149}},\ \bibinfo {pages} {39001} (\bibinfo {year} {2025})},\ \Eprint {https://arxiv.org/abs/2405.20386} {arXiv:2405.20386 [hep-th]} \BibitemShut {NoStop}%
\bibitem [{\citenamefont {Pastor-Guti\'errez}\ \emph {et~al.}(2023)\citenamefont {Pastor-Guti\'errez}, \citenamefont {Pawlowski},\ and\ \citenamefont {Reichert}}]{Pastor-Gutierrez:2022nki}%
  \BibitemOpen
  \bibfield  {author} {\bibinfo {author} {\bibfnamefont {A.}~\bibnamefont {Pastor-Guti\'errez}}, \bibinfo {author} {\bibfnamefont {J.~M.}\ \bibnamefont {Pawlowski}},\ and\ \bibinfo {author} {\bibfnamefont {M.}~\bibnamefont {Reichert}},\ }\bibfield  {title} {\bibinfo {title} {{The Asymptotically Safe Standard Model: From quantum gravity to dynamical chiral symmetry breaking}},\ }\href {https://doi.org/10.21468/SciPostPhys.15.3.105} {\bibfield  {journal} {\bibinfo  {journal} {SciPost Phys.}\ }\textbf {\bibinfo {volume} {15}},\ \bibinfo {pages} {105} (\bibinfo {year} {2023})},\ \Eprint {https://arxiv.org/abs/2207.09817} {arXiv:2207.09817 [hep-th]} \BibitemShut {NoStop}%
\bibitem [{\citenamefont {Heisenberg}(2014)}]{Heisenberg:2014rta}%
  \BibitemOpen
  \bibfield  {author} {\bibinfo {author} {\bibfnamefont {L.}~\bibnamefont {Heisenberg}},\ }\bibfield  {title} {\bibinfo {title} {{Generalization of the Proca Action}},\ }\href {https://doi.org/10.1088/1475-7516/2014/05/015} {\bibfield  {journal} {\bibinfo  {journal} {JCAP}\ }\textbf {\bibinfo {volume} {05}},\ \bibinfo {pages} {015}},\ \Eprint {https://arxiv.org/abs/1402.7026} {arXiv:1402.7026 [hep-th]} \BibitemShut {NoStop}%
\bibitem [{\citenamefont {De~Felice}\ \emph {et~al.}(2016{\natexlab{a}})\citenamefont {De~Felice}, \citenamefont {Heisenberg}, \citenamefont {Kase}, \citenamefont {Mukohyama}, \citenamefont {Tsujikawa},\ and\ \citenamefont {Zhang}}]{DeFelice:2016yws}%
  \BibitemOpen
  \bibfield  {author} {\bibinfo {author} {\bibfnamefont {A.}~\bibnamefont {De~Felice}}, \bibinfo {author} {\bibfnamefont {L.}~\bibnamefont {Heisenberg}}, \bibinfo {author} {\bibfnamefont {R.}~\bibnamefont {Kase}}, \bibinfo {author} {\bibfnamefont {S.}~\bibnamefont {Mukohyama}}, \bibinfo {author} {\bibfnamefont {S.}~\bibnamefont {Tsujikawa}},\ and\ \bibinfo {author} {\bibfnamefont {Y.-l.}\ \bibnamefont {Zhang}},\ }\bibfield  {title} {\bibinfo {title} {{Cosmology in generalized Proca theories}},\ }\href {https://doi.org/10.1088/1475-7516/2016/06/048} {\bibfield  {journal} {\bibinfo  {journal} {JCAP}\ }\textbf {\bibinfo {volume} {06}},\ \bibinfo {pages} {048}},\ \Eprint {https://arxiv.org/abs/1603.05806} {arXiv:1603.05806 [gr-qc]} \BibitemShut {NoStop}%
\bibitem [{\citenamefont {De~Felice}\ \emph {et~al.}(2016{\natexlab{b}})\citenamefont {De~Felice}, \citenamefont {Heisenberg}, \citenamefont {Kase}, \citenamefont {Mukohyama}, \citenamefont {Tsujikawa},\ and\ \citenamefont {Zhang}}]{DeFelice:2016uil}%
  \BibitemOpen
  \bibfield  {author} {\bibinfo {author} {\bibfnamefont {A.}~\bibnamefont {De~Felice}}, \bibinfo {author} {\bibfnamefont {L.}~\bibnamefont {Heisenberg}}, \bibinfo {author} {\bibfnamefont {R.}~\bibnamefont {Kase}}, \bibinfo {author} {\bibfnamefont {S.}~\bibnamefont {Mukohyama}}, \bibinfo {author} {\bibfnamefont {S.}~\bibnamefont {Tsujikawa}},\ and\ \bibinfo {author} {\bibfnamefont {Y.-l.}\ \bibnamefont {Zhang}},\ }\bibfield  {title} {\bibinfo {title} {{Effective gravitational couplings for cosmological perturbations in generalized Proca theories}},\ }\href {https://doi.org/10.1103/PhysRevD.94.044024} {\bibfield  {journal} {\bibinfo  {journal} {Phys. Rev. D}\ }\textbf {\bibinfo {volume} {94}},\ \bibinfo {pages} {044024} (\bibinfo {year} {2016}{\natexlab{b}})},\ \Eprint {https://arxiv.org/abs/1605.05066} {arXiv:1605.05066 [gr-qc]} \BibitemShut {NoStop}%
\bibitem [{\citenamefont {de~Felice}\ \emph {et~al.}(2017)\citenamefont {de~Felice}, \citenamefont {Heisenberg},\ and\ \citenamefont {Tsujikawa}}]{deFelice:2017paw}%
  \BibitemOpen
  \bibfield  {author} {\bibinfo {author} {\bibfnamefont {A.}~\bibnamefont {de~Felice}}, \bibinfo {author} {\bibfnamefont {L.}~\bibnamefont {Heisenberg}},\ and\ \bibinfo {author} {\bibfnamefont {S.}~\bibnamefont {Tsujikawa}},\ }\bibfield  {title} {\bibinfo {title} {{Observational constraints on generalized Proca theories}},\ }\href {https://doi.org/10.1103/PhysRevD.95.123540} {\bibfield  {journal} {\bibinfo  {journal} {Phys. Rev. D}\ }\textbf {\bibinfo {volume} {95}},\ \bibinfo {pages} {123540} (\bibinfo {year} {2017})},\ \Eprint {https://arxiv.org/abs/1703.09573} {arXiv:1703.09573 [astro-ph.CO]} \BibitemShut {NoStop}%
\bibitem [{\citenamefont {Heisenberg}\ \emph {et~al.}(2016)\citenamefont {Heisenberg}, \citenamefont {Kase},\ and\ \citenamefont {Tsujikawa}}]{Heisenberg:2016eld}%
  \BibitemOpen
  \bibfield  {author} {\bibinfo {author} {\bibfnamefont {L.}~\bibnamefont {Heisenberg}}, \bibinfo {author} {\bibfnamefont {R.}~\bibnamefont {Kase}},\ and\ \bibinfo {author} {\bibfnamefont {S.}~\bibnamefont {Tsujikawa}},\ }\bibfield  {title} {\bibinfo {title} {{Beyond generalized Proca theories}},\ }\href {https://doi.org/10.1016/j.physletb.2016.07.052} {\bibfield  {journal} {\bibinfo  {journal} {Phys. Lett. B}\ }\textbf {\bibinfo {volume} {760}},\ \bibinfo {pages} {617} (\bibinfo {year} {2016})},\ \Eprint {https://arxiv.org/abs/1605.05565} {arXiv:1605.05565 [hep-th]} \BibitemShut {NoStop}%
\bibitem [{\citenamefont {De~Felice}\ \emph {et~al.}(2016{\natexlab{c}})\citenamefont {De~Felice}, \citenamefont {Heisenberg}, \citenamefont {Kase}, \citenamefont {Tsujikawa}, \citenamefont {Zhang},\ and\ \citenamefont {Zhao}}]{DeFelice:2016cri}%
  \BibitemOpen
  \bibfield  {author} {\bibinfo {author} {\bibfnamefont {A.}~\bibnamefont {De~Felice}}, \bibinfo {author} {\bibfnamefont {L.}~\bibnamefont {Heisenberg}}, \bibinfo {author} {\bibfnamefont {R.}~\bibnamefont {Kase}}, \bibinfo {author} {\bibfnamefont {S.}~\bibnamefont {Tsujikawa}}, \bibinfo {author} {\bibfnamefont {Y.-l.}\ \bibnamefont {Zhang}},\ and\ \bibinfo {author} {\bibfnamefont {G.-B.}\ \bibnamefont {Zhao}},\ }\bibfield  {title} {\bibinfo {title} {{Screening fifth forces in generalized Proca theories}},\ }\href {https://doi.org/10.1103/PhysRevD.93.104016} {\bibfield  {journal} {\bibinfo  {journal} {Phys. Rev. D}\ }\textbf {\bibinfo {volume} {93}},\ \bibinfo {pages} {104016} (\bibinfo {year} {2016}{\natexlab{c}})},\ \Eprint {https://arxiv.org/abs/1602.00371} {arXiv:1602.00371 [gr-qc]} \BibitemShut {NoStop}%
\bibitem [{\citenamefont {Wetterich}(1993)}]{Wetterich:1992yh}%
  \BibitemOpen
  \bibfield  {author} {\bibinfo {author} {\bibfnamefont {C.}~\bibnamefont {Wetterich}},\ }\bibfield  {title} {\bibinfo {title} {{Exact evolution equation for the effective potential}},\ }\href {https://doi.org/10.1016/0370-2693(93)90726-X} {\bibfield  {journal} {\bibinfo  {journal} {Phys. Lett. B}\ }\textbf {\bibinfo {volume} {301}},\ \bibinfo {pages} {90} (\bibinfo {year} {1993})},\ \Eprint {https://arxiv.org/abs/1710.05815} {arXiv:1710.05815 [hep-th]} \BibitemShut {NoStop}%
\bibitem [{\citenamefont {Pawlowski}\ and\ \citenamefont {Reichert}(2021)}]{Pawlowski:2020qer}%
  \BibitemOpen
  \bibfield  {author} {\bibinfo {author} {\bibfnamefont {J.~M.}\ \bibnamefont {Pawlowski}}\ and\ \bibinfo {author} {\bibfnamefont {M.}~\bibnamefont {Reichert}},\ }\bibfield  {title} {\bibinfo {title} {{Quantum Gravity: A Fluctuating Point of View}},\ }\href {https://doi.org/10.3389/fphy.2020.551848} {\bibfield  {journal} {\bibinfo  {journal} {Front. in Phys.}\ }\textbf {\bibinfo {volume} {8}},\ \bibinfo {pages} {551848} (\bibinfo {year} {2021})},\ \Eprint {https://arxiv.org/abs/2007.10353} {arXiv:2007.10353 [hep-th]} \BibitemShut {NoStop}%
\bibitem [{\citenamefont {Denz}\ \emph {et~al.}(2018)\citenamefont {Denz}, \citenamefont {Pawlowski},\ and\ \citenamefont {Reichert}}]{Denz:2016qks}%
  \BibitemOpen
  \bibfield  {author} {\bibinfo {author} {\bibfnamefont {T.}~\bibnamefont {Denz}}, \bibinfo {author} {\bibfnamefont {J.~M.}\ \bibnamefont {Pawlowski}},\ and\ \bibinfo {author} {\bibfnamefont {M.}~\bibnamefont {Reichert}},\ }\bibfield  {title} {\bibinfo {title} {{Towards apparent convergence in asymptotically safe quantum gravity}},\ }\href {https://doi.org/10.1140/epjc/s10052-018-5806-0} {\bibfield  {journal} {\bibinfo  {journal} {Eur. Phys. J. C}\ }\textbf {\bibinfo {volume} {78}},\ \bibinfo {pages} {336} (\bibinfo {year} {2018})},\ \Eprint {https://arxiv.org/abs/1612.07315} {arXiv:1612.07315 [hep-th]} \BibitemShut {NoStop}%
\bibitem [{\citenamefont {Eichhorn}\ \emph {et~al.}(2018)\citenamefont {Eichhorn}, \citenamefont {Labus}, \citenamefont {Pawlowski},\ and\ \citenamefont {Reichert}}]{Eichhorn:2018akn}%
  \BibitemOpen
  \bibfield  {author} {\bibinfo {author} {\bibfnamefont {A.}~\bibnamefont {Eichhorn}}, \bibinfo {author} {\bibfnamefont {P.}~\bibnamefont {Labus}}, \bibinfo {author} {\bibfnamefont {J.~M.}\ \bibnamefont {Pawlowski}},\ and\ \bibinfo {author} {\bibfnamefont {M.}~\bibnamefont {Reichert}},\ }\bibfield  {title} {\bibinfo {title} {{Effective universality in quantum gravity}},\ }\href {https://doi.org/10.21468/SciPostPhys.5.4.031} {\bibfield  {journal} {\bibinfo  {journal} {SciPost Phys.}\ }\textbf {\bibinfo {volume} {5}},\ \bibinfo {pages} {031} (\bibinfo {year} {2018})},\ \Eprint {https://arxiv.org/abs/1804.00012} {arXiv:1804.00012 [hep-th]} \BibitemShut {NoStop}%
\bibitem [{\citenamefont {Eichhorn}\ \emph {et~al.}(2019)\citenamefont {Eichhorn}, \citenamefont {Lippoldt}, \citenamefont {Pawlowski}, \citenamefont {Reichert},\ and\ \citenamefont {Schiffer}}]{Eichhorn:2018ydy}%
  \BibitemOpen
  \bibfield  {author} {\bibinfo {author} {\bibfnamefont {A.}~\bibnamefont {Eichhorn}}, \bibinfo {author} {\bibfnamefont {S.}~\bibnamefont {Lippoldt}}, \bibinfo {author} {\bibfnamefont {J.~M.}\ \bibnamefont {Pawlowski}}, \bibinfo {author} {\bibfnamefont {M.}~\bibnamefont {Reichert}},\ and\ \bibinfo {author} {\bibfnamefont {M.}~\bibnamefont {Schiffer}},\ }\bibfield  {title} {\bibinfo {title} {{How perturbative is quantum gravity?}},\ }\href {https://doi.org/10.1016/j.physletb.2019.01.071} {\bibfield  {journal} {\bibinfo  {journal} {Phys. Lett. B}\ }\textbf {\bibinfo {volume} {792}},\ \bibinfo {pages} {310} (\bibinfo {year} {2019})},\ \Eprint {https://arxiv.org/abs/1810.02828} {arXiv:1810.02828 [hep-th]} \BibitemShut {NoStop}%
\bibitem [{\citenamefont {Jung}\ and\ \citenamefont {von Smekal}(2019)}]{Jung:2019nnr}%
  \BibitemOpen
  \bibfield  {author} {\bibinfo {author} {\bibfnamefont {C.}~\bibnamefont {Jung}}\ and\ \bibinfo {author} {\bibfnamefont {L.}~\bibnamefont {von Smekal}},\ }\bibfield  {title} {\bibinfo {title} {{Fluctuating vector mesons in analytically continued functional RG flow equations}},\ }\href {https://doi.org/10.1103/PhysRevD.100.116009} {\bibfield  {journal} {\bibinfo  {journal} {Phys. Rev. D}\ }\textbf {\bibinfo {volume} {100}},\ \bibinfo {pages} {116009} (\bibinfo {year} {2019})},\ \Eprint {https://arxiv.org/abs/1909.13712} {arXiv:1909.13712 [hep-ph]} \BibitemShut {NoStop}%
\bibitem [{\citenamefont {Litim}\ and\ \citenamefont {Pawlowski}(1998)}]{Litim:1998qi}%
  \BibitemOpen
  \bibfield  {author} {\bibinfo {author} {\bibfnamefont {D.~F.}\ \bibnamefont {Litim}}\ and\ \bibinfo {author} {\bibfnamefont {J.~M.}\ \bibnamefont {Pawlowski}},\ }\bibfield  {title} {\bibinfo {title} {{Flow equations for Yang-Mills theories in general axial gauges}},\ }\href {https://doi.org/10.1016/S0370-2693(98)00761-8} {\bibfield  {journal} {\bibinfo  {journal} {Phys.Lett.}\ }\textbf {\bibinfo {volume} {B435}},\ \bibinfo {pages} {181} (\bibinfo {year} {1998})},\ \Eprint {https://arxiv.org/abs/hep-th/9802064} {arXiv:hep-th/9802064 [hep-th]} \BibitemShut {NoStop}%
\bibitem [{\citenamefont {Heisenberg}\ \emph {et~al.}(2026)\citenamefont {Heisenberg}, \citenamefont {Platania},\ and\ \citenamefont {Rufrano~Aliberti}}]{Heisenberg:2026rdk}%
  \BibitemOpen
  \bibfield  {author} {\bibinfo {author} {\bibfnamefont {L.}~\bibnamefont {Heisenberg}}, \bibinfo {author} {\bibfnamefont {A.}~\bibnamefont {Platania}},\ and\ \bibinfo {author} {\bibfnamefont {S.}~\bibnamefont {Rufrano~Aliberti}},\ }\bibfield  {title} {\bibinfo {title} {{Asymptotic Safety in Generalized Proca Theories}},\ }\href@noop {} {\  (\bibinfo {year} {2026})},\ \Eprint {https://arxiv.org/abs/2601.20944} {arXiv:2601.20944 [hep-th]} \BibitemShut {NoStop}%
\bibitem [{\citenamefont {Meibohm}\ \emph {et~al.}(2016)\citenamefont {Meibohm}, \citenamefont {Pawlowski},\ and\ \citenamefont {Reichert}}]{Meibohm:2015twa}%
  \BibitemOpen
  \bibfield  {author} {\bibinfo {author} {\bibfnamefont {J.}~\bibnamefont {Meibohm}}, \bibinfo {author} {\bibfnamefont {J.~M.}\ \bibnamefont {Pawlowski}},\ and\ \bibinfo {author} {\bibfnamefont {M.}~\bibnamefont {Reichert}},\ }\bibfield  {title} {\bibinfo {title} {{Asymptotic safety of gravity-matter systems}},\ }\href {https://doi.org/10.1103/PhysRevD.93.084035} {\bibfield  {journal} {\bibinfo  {journal} {Phys. Rev. D}\ }\textbf {\bibinfo {volume} {93}},\ \bibinfo {pages} {084035} (\bibinfo {year} {2016})},\ \Eprint {https://arxiv.org/abs/1510.07018} {arXiv:1510.07018 [hep-th]} \BibitemShut {NoStop}%
\bibitem [{\citenamefont {Christiansen}\ \emph {et~al.}(2018)\citenamefont {Christiansen}, \citenamefont {Litim}, \citenamefont {Pawlowski},\ and\ \citenamefont {Reichert}}]{Christiansen:2017cxa}%
  \BibitemOpen
  \bibfield  {author} {\bibinfo {author} {\bibfnamefont {N.}~\bibnamefont {Christiansen}}, \bibinfo {author} {\bibfnamefont {D.~F.}\ \bibnamefont {Litim}}, \bibinfo {author} {\bibfnamefont {J.~M.}\ \bibnamefont {Pawlowski}},\ and\ \bibinfo {author} {\bibfnamefont {M.}~\bibnamefont {Reichert}},\ }\bibfield  {title} {\bibinfo {title} {{Asymptotic safety of gravity with matter}},\ }\href {https://doi.org/10.1103/PhysRevD.97.106012} {\bibfield  {journal} {\bibinfo  {journal} {Phys. Rev. D}\ }\textbf {\bibinfo {volume} {97}},\ \bibinfo {pages} {106012} (\bibinfo {year} {2018})},\ \Eprint {https://arxiv.org/abs/1710.04669} {arXiv:1710.04669 [hep-th]} \BibitemShut {NoStop}%
\bibitem [{\citenamefont {Denz}\ \emph {et~al.}()\citenamefont {Denz}, \citenamefont {Held}, \citenamefont {Pawlowski}, \citenamefont {Reichert},\ and\ \citenamefont {Rodigast}}]{vertexpand}%
  \BibitemOpen
  \bibfield  {author} {\bibinfo {author} {\bibfnamefont {T.}~\bibnamefont {Denz}}, \bibinfo {author} {\bibfnamefont {A.}~\bibnamefont {Held}}, \bibinfo {author} {\bibfnamefont {J.~M.}\ \bibnamefont {Pawlowski}}, \bibinfo {author} {\bibfnamefont {M.}~\bibnamefont {Reichert}},\ and\ \bibinfo {author} {\bibfnamefont {A.}~\bibnamefont {Rodigast}},\ }\href@noop {} {\bibinfo  {journal} {in preparation}\ }\BibitemShut {NoStop}%
\bibitem [{\citenamefont {Huber}\ \emph {et~al.}(2020)\citenamefont {Huber}, \citenamefont {Cyrol},\ and\ \citenamefont {Pawlowski}}]{Huber:2019dkb}%
  \BibitemOpen
\bibfield  {journal} {  }\bibfield  {author} {\bibinfo {author} {\bibfnamefont {M.~Q.}\ \bibnamefont {Huber}}, \bibinfo {author} {\bibfnamefont {A.~K.}\ \bibnamefont {Cyrol}},\ and\ \bibinfo {author} {\bibfnamefont {J.~M.}\ \bibnamefont {Pawlowski}},\ }\bibfield  {title} {\bibinfo {title} {{DoFun 3.0: Functional equations in Mathematica}},\ }\href {https://doi.org/10.1016/j.cpc.2019.107058} {\bibfield  {journal} {\bibinfo  {journal} {Comput. Phys. Commun.}\ }\textbf {\bibinfo {volume} {248}},\ \bibinfo {pages} {107058} (\bibinfo {year} {2020})},\ \Eprint {https://arxiv.org/abs/1908.02760} {arXiv:1908.02760 [hep-ph]} \BibitemShut {NoStop}%
\bibitem [{\citenamefont {Cyrol}\ \emph {et~al.}(2017)\citenamefont {Cyrol}, \citenamefont {Mitter},\ and\ \citenamefont {Strodthoff}}]{Cyrol:2016zqb}%
  \BibitemOpen
  \bibfield  {author} {\bibinfo {author} {\bibfnamefont {A.~K.}\ \bibnamefont {Cyrol}}, \bibinfo {author} {\bibfnamefont {M.}~\bibnamefont {Mitter}},\ and\ \bibinfo {author} {\bibfnamefont {N.}~\bibnamefont {Strodthoff}},\ }\bibfield  {title} {\bibinfo {title} {{FormTracer - A Mathematica Tracing Package Using FORM}},\ }\href {https://doi.org/10.1016/j.cpc.2017.05.024} {\bibfield  {journal} {\bibinfo  {journal} {Comput. Phys. Commun.}\ }\textbf {\bibinfo {volume} {219}},\ \bibinfo {pages} {346} (\bibinfo {year} {2017})},\ \Eprint {https://arxiv.org/abs/1610.09331} {arXiv:1610.09331 [hep-ph]} \BibitemShut {NoStop}%
\bibitem [{\citenamefont {Vermaseren}(2000)}]{Vermaseren:2000nd}%
  \BibitemOpen
  \bibfield  {author} {\bibinfo {author} {\bibfnamefont {J.~A.~M.}\ \bibnamefont {Vermaseren}},\ }\bibfield  {title} {\bibinfo {title} {{New features of FORM}},\ }\href@noop {} {\  (\bibinfo {year} {2000})},\ \Eprint {https://arxiv.org/abs/math-ph/0010025} {arXiv:math-ph/0010025} \BibitemShut {NoStop}%
\bibitem [{\citenamefont {Kuipers}\ \emph {et~al.}(2013)\citenamefont {Kuipers}, \citenamefont {Ueda}, \citenamefont {Vermaseren},\ and\ \citenamefont {Vollinga}}]{Kuipers:2012rf}%
  \BibitemOpen
  \bibfield  {author} {\bibinfo {author} {\bibfnamefont {J.}~\bibnamefont {Kuipers}}, \bibinfo {author} {\bibfnamefont {T.}~\bibnamefont {Ueda}}, \bibinfo {author} {\bibfnamefont {J.~A.~M.}\ \bibnamefont {Vermaseren}},\ and\ \bibinfo {author} {\bibfnamefont {J.}~\bibnamefont {Vollinga}},\ }\bibfield  {title} {\bibinfo {title} {{FORM version 4.0}},\ }\href {https://doi.org/10.1016/j.cpc.2012.12.028} {\bibfield  {journal} {\bibinfo  {journal} {Comput. Phys. Commun.}\ }\textbf {\bibinfo {volume} {184}},\ \bibinfo {pages} {1453} (\bibinfo {year} {2013})},\ \Eprint {https://arxiv.org/abs/1203.6543} {arXiv:1203.6543 [cs.SC]} \BibitemShut {NoStop}%
\bibitem [{\citenamefont {Ruijl}\ \emph {et~al.}(2017)\citenamefont {Ruijl}, \citenamefont {Ueda},\ and\ \citenamefont {Vermaseren}}]{Ruijl:2017dtg}%
  \BibitemOpen
  \bibfield  {author} {\bibinfo {author} {\bibfnamefont {B.}~\bibnamefont {Ruijl}}, \bibinfo {author} {\bibfnamefont {T.}~\bibnamefont {Ueda}},\ and\ \bibinfo {author} {\bibfnamefont {J.}~\bibnamefont {Vermaseren}},\ }\bibfield  {title} {\bibinfo {title} {{FORM version 4.2}},\ }\href@noop {} {\  (\bibinfo {year} {2017})},\ \Eprint {https://arxiv.org/abs/1707.06453} {arXiv:1707.06453 [hep-ph]} \BibitemShut {NoStop}%
\bibitem [{\citenamefont {Davies}\ \emph {et~al.}(2026)\citenamefont {Davies}, \citenamefont {Kaneko}, \citenamefont {Marinissen}, \citenamefont {Ueda},\ and\ \citenamefont {Vermaseren}}]{Davies:2026cci}%
  \BibitemOpen
  \bibfield  {author} {\bibinfo {author} {\bibfnamefont {J.}~\bibnamefont {Davies}}, \bibinfo {author} {\bibfnamefont {T.}~\bibnamefont {Kaneko}}, \bibinfo {author} {\bibfnamefont {C.}~\bibnamefont {Marinissen}}, \bibinfo {author} {\bibfnamefont {T.}~\bibnamefont {Ueda}},\ and\ \bibinfo {author} {\bibfnamefont {J.~A.~M.}\ \bibnamefont {Vermaseren}},\ }\bibfield  {title} {\bibinfo {title} {{FORM Version 5.0}},\ }\href@noop {} {\  (\bibinfo {year} {2026})},\ \Eprint {https://arxiv.org/abs/2601.19982} {arXiv:2601.19982 [hep-ph]} \BibitemShut {NoStop}%
\bibitem [{\citenamefont {\href{https://as-codebase.quantum-spacetime.net/}{Asymptotic Safety Codebase}}(2024)}]{AScodebase}%
  \BibitemOpen
  \bibfield  {author} {\bibinfo {author} {\bibnamefont {\href{https://as-codebase.quantum-spacetime.net/}{Asymptotic Safety Codebase}}},\ }\href@noop {} {} (\bibinfo {year} {2024})\BibitemShut {NoStop}%
\bibitem [{\citenamefont {Litim}(2001)}]{Litim:2001up}%
  \BibitemOpen
  \bibfield  {author} {\bibinfo {author} {\bibfnamefont {D.~F.}\ \bibnamefont {Litim}},\ }\bibfield  {title} {\bibinfo {title} {{Optimized renormalization group flows}},\ }\href {https://doi.org/10.1103/PhysRevD.64.105007} {\bibfield  {journal} {\bibinfo  {journal} {Phys. Rev. D}\ }\textbf {\bibinfo {volume} {64}},\ \bibinfo {pages} {105007} (\bibinfo {year} {2001})},\ \Eprint {https://arxiv.org/abs/hep-th/0103195} {arXiv:hep-th/0103195} \BibitemShut {NoStop}%
\bibitem [{\citenamefont {Christiansen}\ \emph {et~al.}(2016)\citenamefont {Christiansen}, \citenamefont {Knorr}, \citenamefont {Pawlowski},\ and\ \citenamefont {Rodigast}}]{Christiansen:2014raa}%
  \BibitemOpen
  \bibfield  {author} {\bibinfo {author} {\bibfnamefont {N.}~\bibnamefont {Christiansen}}, \bibinfo {author} {\bibfnamefont {B.}~\bibnamefont {Knorr}}, \bibinfo {author} {\bibfnamefont {J.~M.}\ \bibnamefont {Pawlowski}},\ and\ \bibinfo {author} {\bibfnamefont {A.}~\bibnamefont {Rodigast}},\ }\bibfield  {title} {\bibinfo {title} {{Global Flows in Quantum Gravity}},\ }\href {https://doi.org/10.1103/PhysRevD.93.044036} {\bibfield  {journal} {\bibinfo  {journal} {Phys. Rev. D}\ }\textbf {\bibinfo {volume} {93}},\ \bibinfo {pages} {044036} (\bibinfo {year} {2016})},\ \Eprint {https://arxiv.org/abs/1403.1232} {arXiv:1403.1232 [hep-th]} \BibitemShut {NoStop}%
\bibitem [{\citenamefont {Christiansen}\ \emph {et~al.}(2015)\citenamefont {Christiansen}, \citenamefont {Knorr}, \citenamefont {Meibohm}, \citenamefont {Pawlowski},\ and\ \citenamefont {Reichert}}]{Christiansen:2015rva}%
  \BibitemOpen
  \bibfield  {author} {\bibinfo {author} {\bibfnamefont {N.}~\bibnamefont {Christiansen}}, \bibinfo {author} {\bibfnamefont {B.}~\bibnamefont {Knorr}}, \bibinfo {author} {\bibfnamefont {J.}~\bibnamefont {Meibohm}}, \bibinfo {author} {\bibfnamefont {J.~M.}\ \bibnamefont {Pawlowski}},\ and\ \bibinfo {author} {\bibfnamefont {M.}~\bibnamefont {Reichert}},\ }\bibfield  {title} {\bibinfo {title} {{Local Quantum Gravity}},\ }\href {https://doi.org/10.1103/PhysRevD.92.121501} {\bibfield  {journal} {\bibinfo  {journal} {Phys. Rev. D}\ }\textbf {\bibinfo {volume} {92}},\ \bibinfo {pages} {121501} (\bibinfo {year} {2015})},\ \Eprint {https://arxiv.org/abs/1506.07016} {arXiv:1506.07016 [hep-th]} \BibitemShut {NoStop}%
\end{thebibliography}%

\end{document}